\documentclass[superscriptaddress,twocolumn,amsmath,amssymb]{revtex4-1}
\pdfoutput=1
\usepackage{graphicx}
\usepackage{eurosym}
\usepackage{amsmath}
\usepackage{amssymb}
\usepackage{dcolumn}
\usepackage{bm}
\usepackage{subfigure}
\usepackage{soul}
\usepackage{algcompatible}
\usepackage{newfloat}
\usepackage{comment}
\DeclareFloatingEnvironment[
    fileext=loa,
    listname=List of Algorithms,
    name=ALGORITHM,
    placement=tbhp,
]{algorithm}

\def\avg#1{\mathinner{\langle{#1}\rangle}}
\def\bra#1{\mathinner{\langle{#1}|}}
\def\ket#1{\mathinner{|{#1}\rangle}}
\newcommand{\braket}[2]{\langle #1|#2\rangle}

\DeclareMathOperator{\Tr}{Tr}
\DeclareMathOperator{\Var}{Var}

\newcommand{\Conv}{\mathop{\scalebox{1.5}{\raisebox{-0.2ex}{$\ast$}}}}%

\newcommand{\ignore}[1]{}

\newcommand{\be}{\begin{equaArXivtion}}
\newcommand{\ee}{\end{equation}}
\newcommand{\ba}{\begin{eqnarray}}
\newcommand{\ea}{\end{eqnarray}}

\begin{document}

\title{The theory of variational hybrid quantum-classical algorithms}

\author{Jarrod R. McClean} 
\email[Corresponding author: ]{jmcclean@lbl.gov}
\affiliation{Computational Research Division, Lawrence Berkeley National Laboratory, Berkeley, CA 94720, USA}
\author{Jonathan Romero}
\affiliation{Department of Chemistry and Chemical Biology, Harvard University, Cambridge, MA 02138, USA}
\author{Ryan Babbush} 
\affiliation{Google, Venice, CA 90291, USA}
\author{Al\'an Aspuru-Guzik}
\email[Corresponding author: ]{aspuru@chemistry.harvard.edu}
\affiliation{Department of Chemistry and Chemical Biology, Harvard University, Cambridge, MA 02138, USA}

\begin{abstract} 
Many quantum algorithms have daunting resource requirements when compared to what is available today.  To address this discrepancy, a quantum-classical hybrid optimization scheme known as ``the quantum variational eigensolver'' was developed~\cite{Peruzzo2014} with the philosophy that even minimal quantum resources could be made useful when used in conjunction with classical routines. In this work we extend the general theory of this algorithm and suggest algorithmic improvements for practical implementations.  Specifically, we develop a variational adiabatic ansatz and explore unitary coupled cluster where we establish a connection from second order unitary coupled cluster to universal gate sets through relaxation of exponential splitting.  We introduce the concept of quantum variational error suppression that allows some errors to be suppressed naturally in this algorithm on a pre-threshold quantum device.  Additionally, we analyze truncation and correlated sampling in Hamiltonian averaging as ways to reduce the cost of this procedure. Finally, we show how the use of modern derivative free optimization techniques can offer dramatic computational savings of up to three orders of magnitude over previously used optimization techniques.
\end{abstract}

\maketitle

\section{Introduction}
Eigenvalue and more general optimization problems lie at the heart of applications and technologies ranging from Google's Page Rank and aircraft design to quantum simulation and quantum chemistry~\cite{Page:1999,Cullum_1994,Golub_2000}.  Quantum computers promise to provide ground breaking advances in our ability to solve these problems by offering solutions that may be exponentially faster than the classical equivalent in some cases.  However, delivering on these promises may require overcoming considerable technological challenges.

Since the initial proposal by Richard Feynman~\cite{Feynman1982}, a number of advances have been made in understanding how to use a quantum computer to help solve eigenvalue and optimization problems.  The quantum simulation algorithms of Abrams and Lloyd~\cite{Abrams1997,Abrams1999} showed how eigenvalues corresponding to some Hermitian operator could be extracted from eigenvectors exponentially faster with respect to dimension than the classical equivalent.  Leveraging this idea, Aspuru-Guzik et. al. showed how one could perform exact quantum chemistry computations in polynomial time for some instances, pushing the boundaries of predictive quantum chemistry~\cite{Aspuru:2005}.  These ideas have since been tested successfully in proof-of-principle quantum experiments using architectures such as quantum photonics, nitrogen vacancies in diamond, and ion traps~\cite{Lanyon:2010,Lu:2011,Walther:2012,Peruzzo2014,Wang:2015}.  

In recent years, there has been a growing interest in the particular application of quantum chemistry on quantum computers.  As a result, a number of efforts have been made to study the scaling and performance of various algorithms while simultaneously offering dramatic algorithmic improvements~\cite{Wang:2008,Whitfield:2011,Kassal:2011,Seeley2012,Wecker2013,Kais:2014,Hastings2014,Poulin2014,McClean:2014,Babbush:2015,Whitfield:2013a,Whitfield:2015a,Babbush:2015a,Babbush:2015b,Veis:2014,Toloui:2013,Trout:2015,Huh:2015}.  The original proposal of quantum chemistry on a quantum computer also introduced the idea of adiabatic state preparation, closely related to general adiabatic quantum computation.  A number of advances in this field as well as extensions of adiabatic computation concepts to more general optimization problems have arisen as well~\cite{Neven:2008,Veis:2014,Babbush:2014a}.

Unfortunately, despite developments in quantum algorithms and optimization of resource requirements, many of the algorithms have hardware requirements far beyond the capability of near-term quantum computers.  Moreover, the overhead of some asymptotically optimal algorithms is such that even the first quantum computers competitive with classical supercomputers may not be able to run them.  To this end, in 2014 Peruzzo and McClean et al. developed the variational quantum eigensolver (VQE), a hybrid quantum-classical algorithm designed to utilize both quantum and classical resources to find variational solutions to eigenvalue and optimization problems not accessible to traditional classical computers~\cite{Peruzzo2014}.  This algorithm was originally implemented and tested on a photonic quantum chip and has since been extended both theoretically and experimentally to ion trap quantum computers~\cite{Yung:2014,Shen:2015}.

The VQE has the notable property that it can run on any quantum device, making it a candidate for exploring the performance of early quantum computers.  Moreover, the algorithm is designed to take advantage of the strengths of a given architecture.  That is, if some gates or quantum operations may be performed with higher fidelity, then the algorithm can leverage these strengths in the design of the quantum hardware ansatz.  Perhaps one of the most interesting features of the algorithm is its ability to variationally suppress some forms of quantum errors, which is discussed later in this work.  This intrinsic robustness to quantum errors in combination with low coherence time requirements has placed this algorithm as a potential candidate for the first to surpass a classical computer, using a pre-threshold quantum device.  Even in the event that some error correction is required to exceed current computational capabilities, this same robustness may translate to requiring minimal error correction resources when compared with other algorithms.

In this work we aim to present the hybrid quantum-classical variational approach in more detail, offering both theoretical and practical exposition on developments since the original hybrid quantum-classical proposal.  Additionally, although a strength of the variational quantum eigensolver is its ability to adapt to the given hardware, this work will be the first to analyze VQE in the abstract, in a way that is completely general to any quantum device. We begin by reviewing background and notation as well as the outline of the variational quantum eigensolver algorithm.  This is followed by a discussion of ansatz states that allow one to explore classically inaccessible regions of Hilbert space, including a variational formulation of adiabatic state preparation and unitary coupled cluster.  We then explore how this approach may be used to variationally suppress certain types of quantum errors.  Following this, we introduce several computational enhancements to the Hamiltonian averaging method for obtaining expectation values, including the truncation of unimportant terms and grouping terms by commutation and covariance. These enhancements are able to considerably reduce the cost of the procedure. Finally we cover aspects of the classical optimization procedure associated with the VQE and show how modern derivative-free optimization technique have the potential to greatly enhance the efficacy of the method.

\section{Background and Notation}
\subsection{General Quantum Systems and the Variational Principle}
Let us consider a quantum system $S$ composed of $N$ qubits which will act as our quantum computer, and a Hamiltonian $H$ of a different system $Q$ that need have no relation to $S$ other than acting on a space of $\leq N$ qubits.  This Hamiltonian could be derived from a physical system such as a collection of interacting spins or the discretization of an interacting electronic system.  Similarly it could come from the encoding of an optimization problem or the problem Hamiltonian in adiabatic quantum computation.  In all of these instances, one is interested in the eigenvectors and eigenvalues, $\ket{\chi_i}$, $\lambda_i$ of the Hamiltonian $H$, and the goal will be to find and study these eigenvectors and eigenvalues using $S$.  

In the VQE approach, the eigenvectors are encoded by a set of parameters that can be used to prepare them on demand when other observables are desired.  We order the eigenvectors by the eigenvalues such that $\lambda_1 \leq \lambda_2 \leq ... \leq \lambda_N$.  Indeed in many cases, the eigenvectors corresponding to the lowest few eigenvalues and their properties are of primary interest.  In physical systems this is because low-energy states play a dominant role in the properties of the system at modest temperatures, and in optimization problems they often encode the optimal solution.

Recall the expectation value of an operator $O$ with respect to a state $\ket{\Psi}$
\begin{align}
\avg{O}_{\ket{\Psi}} = \frac{\bra{\Psi} O \ket{\Psi}}{\braket{\Psi}{\Psi}}.
\end{align}
We will assume normalization of the wavefunction, $\braket{\Psi}{\Psi}=1$, for the remainder of the work, however attention should be paid in the case of leakage errors from the computational basis.  Our attention is restricted to the class of operators whose expectation value can be measured efficiently on $S$ and mapped to $Q$.  A sufficient condition for this property is that operators have a decomposition into a polynomial sum of simple operators as\begin{align}
O = \sum_\alpha h_\alpha O_{\alpha}
\end{align}
where $O$ is an operator than acts on $Q$, $\alpha$ runs over a number of terms polynomial in the size of the system, $h_\alpha$ is a constant coefficient, each $O_\alpha$ has a simple measurement prescription on the system $S$. This will allow for straightforward determination of expectation values of $O$ on $Q$ by weighted summation of projective measurements on the quantum device $S$.  A simple example of this is the decomposition of a Hermitian operator into a sum of tensor products of Pauli operators weighted by constant coefficients.

Consider a set of real valued parameters $\{\theta_i\}$, which we arrange into a vector $\vec{\theta}$, and the Hamiltonian $H$ of $Q$.  If one prepares $S$ into a quantum state depending on these parameters, $\ket{\Psi(\vec{\theta})}$, then the variational theorem of quantum mechanics states that
\begin{align}
\avg{H}_{\ket{\Psi(\vec{\theta})}} \equiv \avg{H}(\vec{\theta}) = \bra{\Psi(\vec{\theta})}H\ket{\Psi(\vec{\theta})} \geq \lambda_1.
\end{align}
As a result, the optimal choice of $\vec{\theta}$ to approximate the ground state (or eigenvector corresponding to the lowest eigenvalue) is the choice which minimizes $\avg{H}(\vec{\theta})$.  Note that the state is normalized for all choices of $\vec{\theta}$ by the unitarity of quantum evolution or trace preservation under quantum operations in state preparation.  The variational principle also extends to other eigenstates.  If one has constructed an ordered orthonormal set of $k$ approximate eigenstates $\{\ket{\tilde \Psi_i}\}_{i=1}^k$, such that $\avg{H}_{\ket{\tilde \Psi_1}} \leq \avg{H}_{\ket{\tilde \Psi_2}} \leq ... \leq \avg{H}_{\ket{\tilde \Psi_k}}$, then 
\begin{align}
\avg{H}_{\ket{\tilde \Psi_i}} \geq \lambda_i \ \forall \ i \in [1,k]
\end{align}
where $\lambda_i$ are the ordered true eigenvalues of the operator $H$.  Thus, repeated application of the variational principle under orthogonality constraints can yield an approximation to as much of the spectrum as desired, incurring additional cost for each eigenvalue.  Alternatively, one can perform a spectral transform to the Hamiltonian and use the ground-state variational principle to find excited states, as in the folded spectrum method~\cite{Wang:1994}.  That is, minimize $\avg{H'}(\vec{\theta})$ where $H'=(H-\gamma I)^2$ and $\gamma$ is some real parameter.  In the transformed Hamiltonian, the ground state corresponds to the eigenvalue in the original Hamiltonian closest to $\gamma$. 

More generally, the state preparation scheme may be influenced by an environment and would be better represented by an ensemble given by a density matrix $\rho(\vec{\theta})$.  In an ideal scenario where the preparation is error free and a pure state is maintained, $\rho(\vec{\theta}) = \ket{\Psi(\vec{\theta})}\bra{\Psi(\vec{\theta})}$.  In the density matrix formalism, the expectation value of an operator $O$ is given by
\begin{align}
\avg{O}_{\rho} = \Tr[\rho O]
\end{align}
and the ground state variational principle on the Hamiltonian $H$ still holds such that for any approximate density matrix $\rho(\vec{\theta})$, and for all choices of $\vec{\theta}$,
\begin{align}
\avg{H}_{\rho(\vec{\theta})}  \equiv \avg{H}(\vec{\theta}) = \Tr[\rho(\vec{\theta}) H] \geq \lambda_1. 
\end{align}
As a result, the optimal choice of $\vec{\theta}$ to approximate the ground state is that which minimizes $\avg{H}_{\rho(\vec{\theta})}$.  The fact that this principle still holds for mixed states has important consequences for the robustness of the method to errors and environmental influence. By finding the set of parameters that minimizes the energy, one is in effect, finding a set of experimental parameters most likely to produce the ground state on the average, potentially affecting a blind purification of the state being produced.  This ability to suppress errors without knowledge of the mechanism will be elaborated upon later in this work.

Another important quantity is the variance of an operator with respect to a state. For an operator $O$ and a general mixed state $\rho$, this is given by 
\begin{align}
\Var[O]_{\rho} &= \left\langle \left( O - \avg{O}_{\rho} \right)^2 \right\rangle_{\rho}  \\
&= \avg{O^2}_{\rho} - \avg{O}_{\rho}^2.
\end{align}
A variational principle on the variance exists as well, and has been used extensively for optimization in the context of quantum Monte Carlo~\cite{Lester:1994}.  Note that for any eigenstate $\ket{\Psi_k}$ of an operator $O$, the variance is 
given by
\begin{align}
\bra{\Psi_k}O^2\ket{\Psi_k} - \bra{\Psi}O\ket{\Psi_k}^2 = (\lambda_k^2) - (\lambda_k)^2 = 0
\end{align}
and for any approximate eigenstate $\ket{\tilde \Psi}$, we have that
\begin{align}
\Var[O]_{\ket{\tilde \Psi}} \geq 0.
\end{align}

\subsection{Fermionic Hamiltonians and Quantum Chemistry}
While the VQE and its principles can be applied to general quantum problems, an application of particular recent interest is that of quantum chemistry and fermionic Hamiltonians.  Given a set of nuclear charges $Z_i$ and a number of electrons, the standard form of the electronic structure problem is to solve for the eigenvectors and eigenvalues of the electronic Hamiltonian $H$, written as
\begin{align}
H &= - \sum_i \frac{\nabla_{R_i}^2 }{2M_i} - \sum_i \frac{\nabla_{r_i}^2}{2} - \sum_{i,j} \frac{Z_i}{|R_i - r_j|} \notag \\
&+ \sum_{i, j > i} \frac{Z_i Z_j}{|R_i - R_j|} + \sum_{i, j>i} \frac{1}{|r_i - r_j|}
\end{align} 
where atomic units have been used, $R_i$ are nuclear positions, $r_i$ electronic positions, and $M_i$ are nuclear masses.  Due to large separations in the nuclear and electronic masses, an excellent approximation to this problem at the time and energy scales of chemical interest is to treat the nuclei as classical point charges under the Born-Oppenheimer approximation with fixed positions $R_i$.  The problem as written is referred to as the first quantized representation of the quantum chemistry problem.  A number of algorithms have been developed for quantum computers to treat the problem directly within this framework~\cite{Kassal:2008,Toloui:2013,Welch:2014}, however the focus in this work will be on the second quantized treatment.

To reach the practical form of the second quantized Hamiltonian, one must project the problem into a finite, orthogonal, spin-orbital basis, of which we will denote members $\varphi_i$, and impose the requirements of fermion anti-symmetry through the fermion creation and annihilation operators $a_i^\dagger$ and $a_i$.  With these steps, the second quantized Hamiltonian takes the form
\begin{align}
H = \sum_{pq} h_{pq} a^{\dagger}_p a_q + \frac{1}{2} \sum_{pqrs} h_{pqrs} a^{\dagger}_p a^{\dagger}_q a_r a_s
\end{align}
with coefficients determined by the spin orbital basis as
\begin{align}
h_{pq} &= \int \ d\sigma \ \varphi_p^*(\sigma) \left(\frac{\nabla_r^2}{2} - \sum_i \frac{Z_i}{|R_i - r|} \right)\varphi_q^*(\sigma)  \\
h_{pqrs} &= \int \ d\sigma_1 \ d\sigma_2 \frac{\varphi_p^*(\sigma_1)\varphi_q^*(\sigma_2) \varphi_s(\sigma_1)\varphi_r(\sigma_2)}{|r_1 - r_2|}
\end{align}
where $\sigma_i$ describes both the spatial position and spin of an electron as $\sigma_i = (r_i, s_i)$.  The operators $a_i^\dagger$ and $a_i$ obey the standard fermion commutation relations as
\begin{align}
\{a_p^\dagger, a_r\} &\equiv a_p^\dagger a_r + a_r a_p^\dagger = \delta_{p,r} \\
\{a_p^\dagger, a_r^\dagger \} &= \{a_p, a_r\} = 0.
\end{align}

A crucial part of solving these problems on quantum computers is the mapping from fermions to qubits.  The two most common mappings under current study are the Jordan-Wigner transformation~\cite{Jordan1928,Somma2002} and the Bravyi-Kitaev transformation~\cite{Bravyi2000,Seeley2012,Tranter:2015}.  In the case of the Jordan-Wigner transformation, the mapping from fermion operators to qubits is 
\begin{align}
a_p^\dagger  &= ( {\prod\nolimits_{m < p} {\sigma _m^z} } ) \sigma _p^+ \\
a_p &= ( {\prod\nolimits_{m < p} {\sigma _m^z} } )\sigma _p^- \\
\sigma^\pm &\equiv \left( {{\sigma ^x} \mp i{\sigma ^y}} \right)/2
\end{align}

\subsection{Reference States}
Many traditional methods for electronic structure involve the concept of a reference state.  A reference state is a product state that is used as a starting point to define a more general quantum state, and can allow for great formal simplification.  Here we will briefly introduce why they are convenient and useful, and then how they are obtained.

An example spin reference $\ket{\Psi_{\text{s-ref}}}$ and fermion reference state $\ket{\Phi_{\text{f-ref}}}$ might be the general product states 
\begin{align}
\ket{\Psi_{\text{s-ref}}} &= \prod_i^N \left( c_i^0 \ket{0} + c_i^1 \ket{1} \right) \\
\ket{\Phi_{\text{f-ref}}} &= \prod_i^N \left(\sum_j^M c_i^j a_j\right) \ket{}
\end{align}
where $\ket{}$ is the fermion vacuum state, $M$ is the number of sites a fermion can occupy, and $N$ is the number of qubits or fermions. Even though these are separable product states, their manipulation theoretically or preparation on a quantum computer can be cumbersome as written.  However, because they are product states, there exist efficient, local unitary basis transformations $U_s \in SU(2)^{\otimes^N}$ and $U_f \in SU(M)$ such that these states can be rotated into a simple form with weight on a single computational basis state.  That is
\begin{align}
U_s \ket{\Psi_{\text{s-ref}}} &= \ket{000...0} \\
U_f \ket{\Phi_{\text{f-ref}}} &= a^\dagger_N a^\dagger_{N-1} ... a^\dagger_1 \ket{ }.
\end{align}
and because the transformations are local, the transformation of the Hamiltonian to the new basis such that the physical problem remains unchanged is also efficient.  In the case of quantum chemistry, this corresponds to a transformation of the integral terms $h_{pq}$ and $h_{pqrs}$, which may be computed in a time O$(M^5)$ exactly.

These new simpler forms of the state have advantages both in theoretical manipulation, and in ease of preparation with quantum resources.  For example, the preparation of the untransformed spin reference state could require at least $O(N)$ local rotations, not including error correction on a quantum device to prepare from a computational basis state, whereas the new reference is simply the computational basis state from which most computations begin.  Here we have traded modest classical effort in transforming the basis of the Hamiltonian for savings in quantum resources.

These reference states are typically obtained from mean field calculations, which are guaranteed to have product states, such as those given above, as solutions.  In chemistry, this procedure is called Hartree-Fock, and the transformation of the state to the simplified form is known as the canonical condition in the solutions of the Hartree-Fock equations, resulting in the canonical molecular orbitals. 

When the problem is well treated by mean-field theory, it can be shown through perturbation theory that the dominant corrections to the mean-field solution are given by quantum states ``close'' to the mean-field solution in the sense of fermion excitations~\cite{Helgaker:2014} or Hamming distance.  This is the origin of the perturbative MP2 method, configuration interaction, and coupled cluster methods~\cite{Bartlett:2007,Helgaker:2014}, which all solve the problem close to a given reference and have been applied to both electronic and frustrated spin-systems~\cite{Gotze:2011}.

In some problems, particularly when correlation is strong, the mean-field description is a poor starting point for the problem.  In this case, one may still use a reference-like formalism, but starting with an entangled state.  These methods are called multi-reference methods in quantum chemistry~\cite{Helgaker:2014,Ladig:1984,Musia:2011}, and carry considerably more theoretical and computational challenges with them.  In this work, we will highlight how the generalization of methods on a quantum computer to the multi-reference case is often more natural than in the classical case.

\subsection{Algorithm Outline}
To use a variational methodology to find approximations to the eigenvalues and eigenvectors of the Hamiltonian in a quantum computer, it is convenient to break the task into three distinct pieces and outline the algorithm very coarsely as
\begin{enumerate}  
\item Prepare the state $\ket{\Psi(\vec{\theta})}$ or $\rho(\vec{\theta})$ on the quantum computer, where $\vec{\theta}$ can be any adjustable experimental or gate parameter.    
\item Measure the expectation value $\avg{H}(\vec{\theta})$  
\item Use a classical non-linear optimizer such as the Nelder-Mead simplex method to determine new values of $\vec{\theta}$ that decrease $\avg{H}(\vec{\theta})$ 
\item Iterate this procedure until convergence in the value of the energy.  The parameters $\vec{\theta}$ at convergence define the desired state. 
\end{enumerate}
In the coming sections we will elaborate on what is known about each of these steps and offer new algorithmic and conceptual improvements.

\section{State parameterization and preparation}
The set of states a quantum computer can easily manipulate that a classical computer cannot is not yet fully understood~\cite{Mora:2005,Gross:2009,Cai:2015}.  Given the set of parameters $\vec{\theta}$, it's clear that in order for a quantum computer to have an advantage, one would like the state $\ket{\Psi(\vec{\theta})}$ to be good at describing the solution of interest, while also difficult to prepare and/or sample from classically using currently known methods.  Here we will first discuss topics relevant to state preparation for all classes of states in the variational quantum eigensolver, independent of any notion of how difficult they are to prepare classically.  We will then discuss some details concerning two classes of states currently believed to be both good at describing systems of interest and difficult to prepare and/or sample from classically, namely adiabatically parameterized states and (multi-reference) unitary coupled cluster states.

\subsection{Error bounds and distributions}
Once a state $\ket{\Psi (\vec{\theta})}$ has been prepared as a function of some set of parameters $\vec{\theta}$, one would like to know how close this state is to the  solution of the problem being solved.  In this work, we will say a measured value $v$ is known to precision $\epsilon$ based on a normal distribution approximation with standard deviation $\epsilon/2$, which is reasonable given that most of our estimates will be derived from sums of random variates with finite variance, which by the central limit will rapidly converge to a normal distribution.

Suppose, for now, that the goal is to know an eigenvalue of $H$ to within a specified precision $\epsilon$.  Let $\lambda_k$ be the eigenvalue of $H$ closest to $\avg{H}(\vec{\theta})$.  Under these assumptions on the eigenvalue the Weinstein inequalities~\cite{Weinstein:1934,MacDonald:1934} hold
\begin{align}
\avg{H}(\vec{\theta}) + \sqrt{\Var(\vec{\theta}}) \geq \lambda_k \geq \avg{H}(\vec{\theta}) - \sqrt{\Var(\vec{\theta}}).
\end{align}
As a result, a sufficient condition is to rigorously achieve the precision requirement $\epsilon$ on the eigenvalue $\lambda_k$ is
\begin{align}
\Var(\vec{\theta}) \leq \frac{\epsilon^2}{4}
\end{align}
where as one approaches an eigenstate, the variance approaches $0$.  When considering only the ground state, one can derive a simple bound on the quality of the state.  More specifically, in the zero variance limit, if $\lambda_1$ has multiplicity $1$, then the eigenstate corresponding to $\lambda_1$ is reproduced as well.  That is, if a bound on the gap to the first eigenstate $\Delta$ is known in addition to the variance, such that $|\lambda_{1} - \lambda_i| \geq \Delta > 0 \ \forall \ i \neq 1$, and $\epsilon/2 < \Delta$, and we decompose the state into its eigenstate representation $\ket{\Psi(\vec{\theta})} = \sum_i c_i(\vec{\theta}) \ket{\chi_i}$ then we can quantify the quality of state preparation as a function of the measured variance
\begin{align}
|\braket{\Psi(\vec{\theta})}{\chi_1}|^2 = |c_1(\vec{\theta})|^2 \geq \frac{\Delta-\sqrt{\Var(\vec\theta)}}{\Delta}.
\end{align}
For general excited states $k$, one may find a similar bound exists based on a measurement of the variance of the operator and a known bound on the gap $\Delta > 0$, such that
\begin{align}
|\braket{\Psi (\vec{\theta})}{\chi_k}|^2 = |c_k(\vec{\theta})|^2 \geq \frac{\gamma - \Var( \vec{\theta})}{\gamma}
\end{align}
where $\gamma = \left(\Delta + \sqrt{\Var( \vec{\theta})} \right)^2$, and both bounds given here are derived in this appendix.
If one has prior knowledge that a single eigenstate dominates the expansion, such that $|c_k(\vec{\theta})|^2 > 0.5$, and a lower bound $0.5 < \alpha \leq |c_k(\vec{\theta})|^2$, then Delos and Blinder ~\cite{Delos:1967} showed through the method of moments that a tighter lower-bound on the eigenvalue is given by
\begin{align}
\lambda_k \geq \avg{H}(\vec{\theta}) - (\frac{1}{\alpha^2} - 1)^{(1/2)} \Var(\vec{\theta}).
\end{align}
These bounds may be used to estimate the absolute accuracy the minimization procedure obtained within the given basis and decide if the eigenvalue has been determined to the desired accuracy and precision or if the state ansatz should be altered to adjust the cost or accuracy of the procedure.

\subsection{Adiabatically parameterized states}
One type of quantum state that can be explored as a parametric ansatz is that produced by adiabatic state state preparation with a variable path. In adiabatic quantum computation~\cite{Farhi:2000,Farhi:2001,Boixo:2010} and adiabatic state preparation~\cite{Aspuru:2005,Veis:2014} one makes use of the adiabatic theorem~\cite{Born:1928}, which states loosely that if one prepares the lowest eigenstate of an initial Hamiltonian $H_i$, by continuously changing the Hamiltonian from $H_i$ to a final problem Hamiltonian $H_p$, one finishes in the lowest eigenstate of $H_f$ if the evolution was slow enough.  In adiabatic computation, slow enough is quantified relative to the minimum eigenvalue gap between the ground and first excited states along the evolution.  While many developments have occurred in the area of adiabatic quantum computation and modifications to the Hamiltonian, perhaps the most commonly considered form of evolution is defined by
\begin{align}
H(s) = A(s)H_i + B(s)H_p 
\end{align}
where $s \in [0,1]$, $A(0)=B(1)=1$, and $A(1)=B(0) = 0$.  The evolution is controlled by continuously changing the parameter $s$ as a function of time $t$.

Consider the set of all paths of $A(s)$ and $B(s)$ from $0$ to $1$ as a function of time $t \in [0,\tau]$, and denote it $F(\tau)$, where $\tau$ is some finite time.  Label one such path as $f \in F(\tau)$.  In a noiseless coherent situation at $0$K, the unitarity of evolution dictates that the final state of the evolution is uniquely determined by the path $f$.  In this situation, we may write the final pure state as a higher-order function of the path $f$, or $\ket{\Psi[f]}$.  Thus any expectation values of the final state may be written as functionals of the path, $\avg{H}[f]$, and by the variational principle
\begin{align}
\avg{H_p}[f] = \bra{\Psi[f]} H_p \ket{\Psi[f]} \geq \lambda_1
\end{align}
such that the optimal path is the path in $F(\tau)$ that minimizes the value of $\avg{H}[f]$.  This functional minimization may be changed into a standard minimization by parameterizing the path $f$ by a set of parameters $\vec{\theta}$, and performing an optimization on the parameters $\vec{\theta}$ that determine the path.   As such, adiabatic state preparation may be considered as an ansatz to be used in the variational hybrid quantum-classical approach, where the state parameters are the shape or nature of the path. The idea of refining the adiabatic path has been used before in the context of local adiabatic evolution~\cite{Roland:2002} with great success.  The idea here is to achieve similar benefits in an entirely black-box manner, guided only by a variational principle and measurements of the final point of the evolution.

As a simple example, consider a linear path in $F(\tau)$ defined by a single parameter $\theta_1$ that controls how quickly the evolution is performed
\begin{align}
A(s) &= \max(1, \theta_1 s)\\
B(s) &= 1 - A(s)
\end{align}
and the parameter $\theta_1$ is restricted by membership in $F(\tau)$ to $1/\tau \leq \theta_1 < \infty$. In the case of an ideal evolution with enough quantum resources such that the evolution is much longer than required by the problem gap, the adiabatic theorem implies that $H(\theta_1)$ is optimal at the extremal point $\theta_1 = 1/\tau$.  Moreover, in the limit that $\tau \rightarrow \infty$, the adiabatic theorem implies that for any finitely gapped problem $F(\tau)$ contains a path that prepares the exact ground state, and even the simplest linear paths, which are a subset of $F(\tau)$, are sufficient to do so.

Within this simple example, it's not immediately clear why one would want the flexibility offered by the variational quantum eigensolver formulation, as one could choose the linear path with $\theta_1$ minimal without the need for any optimization of $\theta_1$.  However, a more realistic situation may be such that $\tau$ is smaller than the required time of evolution dictated by the problem gap, due to technological constraints or simply human time constraints in a hard problem.  It might also be possible that no good estimate of the gap is known, and one must attempt several paths regardless to establish confidence that the evolution is not too fast to impair accuracy.  One should exercise caution in such attempts however, as the probability of success does not necessarily increase monotonically with evolution time, especially when one is far short of the time required by the problem gap or when errors are present~\cite{Bookatz:2014}.  Moreover, it is known that for systems experiencing decoherence or dephasing on the timescale of evolution that the slowest possible evolution is not optimal in preparing the ground state of the final problem Hamiltonian~\cite{Steffan:2003,Aberg:2005,Crosson:2014}.  In all situations, the final density matrix is determined by the parameters of the path, such that $f$ determines a density matrix $\rho[f] = \rho(\vec{\theta})$, and an optimal choice of parameters can be made without detailed knowledge of the gap or errors present in a system by minimizing $\avg{H_p}[f] = \avg{H_p}(\vec{\theta})=\Tr[\rho(\vec{\theta})H_p]$ as a function of $\vec{\theta}$.  

The Hamiltonians may also be generalized to include intermediate operators~\cite{Farhi:2002a,Hofmann:2014,Crosson:2014,Zeng:2015} such as
\begin{align}
H(s) = A(s)H_i + B(s) H_p +  \sum_j C_j(s) H_{j}
\end{align}
where one considers any number of intermediate Hamiltonians $H_j$ and $C_j$ with $C_j(0)=C_j(1)=0$.  The set of paths satisfying these boundary conditions with available intermediate Hamiltonians $\{H_j\}$, $F(\tau, \{H_{j}\})$, offers more flexibility, and again a guiding principle to select parameters defining the optimal paths is given by the variational principle.

\begin{figure}[t!]
  \centering
    \includegraphics[width=8cm]{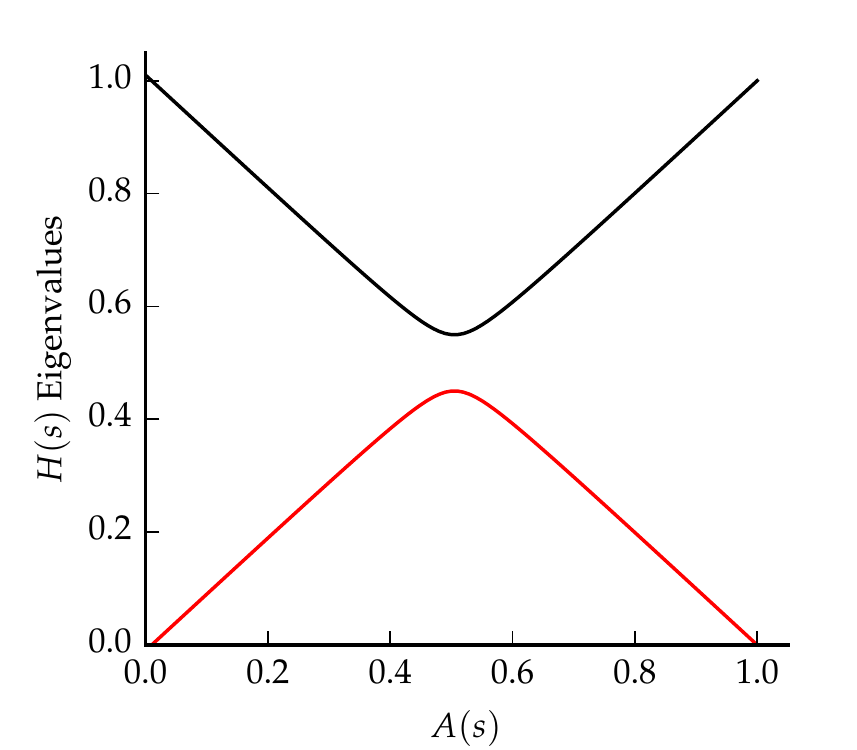}
  \caption{The ground and first excited state eigenvalues of the schedule Hamiltonian $H(s)$ as a function of the annealing path $A(s)$.  This shows the avoided crossing that occurs at $A(s)=1/2$, the size of which is controlled by the perturbation parameters $\epsilon$ in the Hamiltonian, which in our example is set to a value of $\epsilon=0.1$.}
  \label{fig:EigVal}
\end{figure}

From this discussion it is clear that adiabatic state preparation where the path of evolution is defined by some set of parameters $\vec{\theta}$ is one choice of parametric ansatz for the variational quantum eigensolver.  It can be inferred from the known capabilities of adiabatic quantum computation that this ansatz is capable of preparing states that cannot be efficiently prepared or sampled from classically using only a small number of parameters with currently known methods~\cite{Aharonov:2008}.  As seen in the simple linear example, the number of parameters to meet this condition may be as few as $1$ for a linear interpolation that is slow enough in ideal conditions.

\subsubsection{Variational Adiabatic Path Example}
To further illustrate the utility of a variational perspective on adiabatic quantum computational methods in a resource constrained setting, we consider here a simple 1-qubit problem first studied in the adiabatic context in the original work of Farhi et al~\cite{Farhi:2000}.  In particular, we will consider this problem in a resource constrained context where the maximum evolution time $\tau$ is limited. In this problem, the Hamiltonian the initial and problem Hamiltonians are given by
\begin{align}
H_i &= \frac{1}{2} \left(I - \sigma_z\right) + \epsilon \sigma_x \\
H_p &= \frac{1}{2} \left(I + \sigma_z\right).
\end{align}
If we take the following form of the schedule Hamiltonian
\begin{align}
H(s) = A(s) H_i + \left[ 1-A(s) \right] H_p
\end{align}
then the eigenvalues of this problem undergo an avoided crossing with a gap determined by the size of the perturbation $\epsilon$.  For this example we choose $\epsilon=0.1$ and the resulting spectrum is plotted in Fig.\ref{fig:EigVal} as a function of $A(s)$.  Suppose that we are attempting to prepare the ground state of our problem Hamiltonian in a situation where the total evolution time $\tau$ is limited.  

\begin{figure}[t!]
  \centering
    \includegraphics[width=8cm]{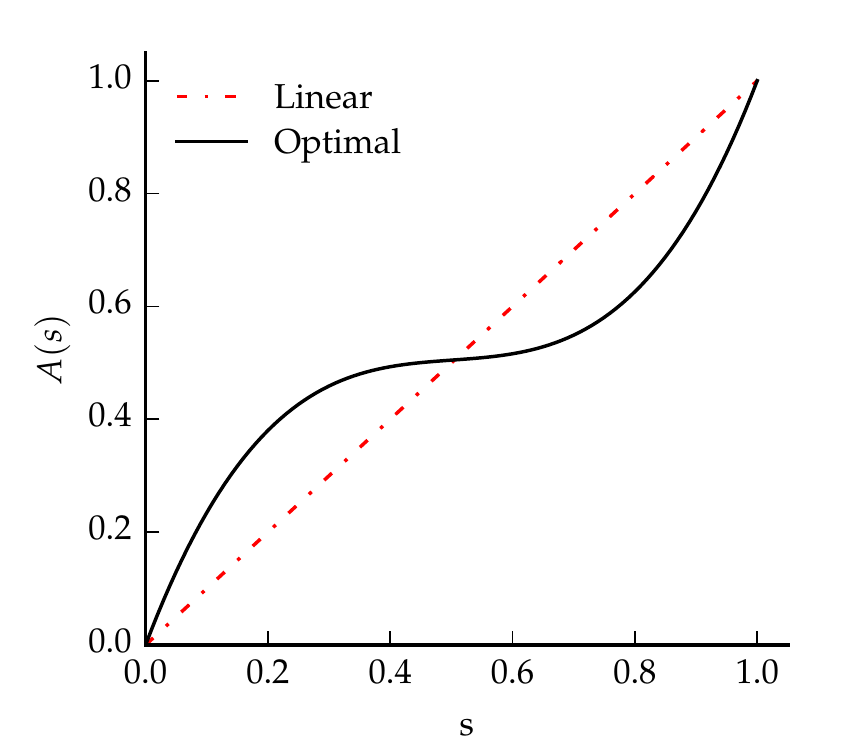}
  \caption{A comparison of the standard linear path $A(s)$ versus the two-parameter split path that is variationally optimal with respect to the expectation value of the Hamiltonian at the final point $H(1)$.  The path naturally slows the evolution near the location of the avoided crossing, but is otherwise only slightly distorted from a standard linear path.}
  \label{fig:Paths}
\end{figure}

We will consider two types of paths, the first of which is a fixed standard linear path as a function of time.  That is $A(s) = s = t/\tau$ with $t \in [0, \tau]$.  The second type of path will be a parameterized path of two variables defined by the best cubic B-spline fit of the 4 points $(0,0), (.15 \tau, \theta_1), (.85\tau, \theta_2), (\tau,1)$, where the the parameters $\theta_i$ are determined by a non-linear minimization the expectation value of the final state in the (possibly non-)adiabatic evolution with fixed maximum evolution time, $\avg{H(1)}(\theta_1, \theta_2)$.  In this simple example we use the Nelder-Mead simplex method to perform a derivative free optimization of $\theta_i$, in analogy to how it might be performed on a quantum device.  We use as an initial condition $\theta_1=.15\tau$ and $\theta_2=.85\tau$ in the optimization, which corresponds to the linear path.

\begin{figure}[t!]
  \centering
    \includegraphics[width=8cm]{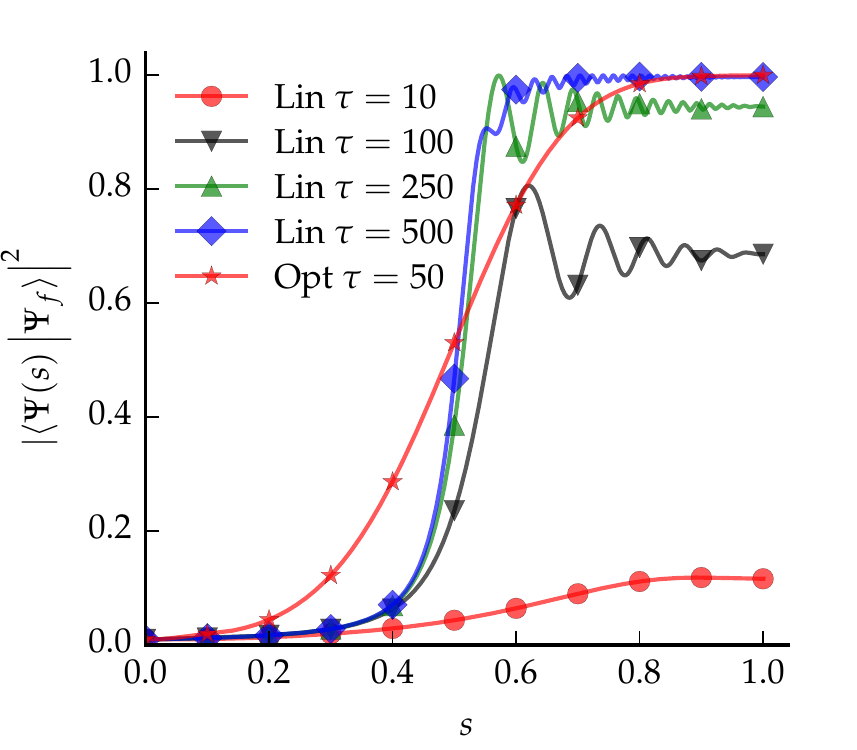}
  \caption{The squared overlap of the system state $\ket{\Psi(s)}$ at parameter value $s$ with the ground state at $H(1)$, $\ket{\Psi_f}$ is show for both the standard Linear (Lin) schedule as well as the variationally optimal spline schedule for different total evolution times $\tau$.  It can be seen here that the performance of the variational schedule offers similar performance to a linear schedule roughly $10$ times as long, indicating an order of magnitude reduction in the quantum evolution time required for the variationally optimal schedule.}
  \label{fig:SuccessProb}
\end{figure}

The resulting variationally optimal adiabatic spline path $A(s)$ is plotted alongside the standard linear path in Fig. \ref{fig:Paths}, which shows that the method naturally finds a path which slows evolution near the closing gap, without any prior knowledge of the spectrum, and only measurements at the endpoint as opposed to the entire path.  The effect of this on the success of preparing  the ground state as a function of the total available evolution time is shown in Fig. \ref{fig:SuccessProb}.  From this figure we observe that the variationally optimal adiabatic spline path is able to achieve similar results to a linear path with roughly $10$ times less evolution time.  That is, at the cost of some classical minimization, we have reduced the quantum evolution time requirement by a factor of $10$ by slightly deforming the schedule in a black-box manner relying only on measurements of the final state of the evolution and no prior knowledge of the problem.  Moreover, even at this reduced evolution time, we achieve the desirable property that the success of the computation is a monotonically increasing function of $s$, which is not true of the linear schedule in this case.

\subsubsection{Pontryagin's Principle and Non-Adiabatic Bang-Bang Quantum Computation}
While adiabatic evolution or attempted adiabatic evolution is one way to prepare a desired state, it is certainly not the only option.  Non-adiabatic evolution opens a different class of potential schedules for preparing a desired state guided by the variational principle.  The form of the schedule Hamiltonian $H(s)$ has a particularly interesting form, namely that it is a linear evolution problem with a control $A(s)$ that effects a linear coupling.  In the theory of optimal control, it is known through application of Pontryagin's minimization principle that the optimal control setting for reaching a desired state of the controlled system when the system has a linear coupling to the control is to have the control at its extremal values~\cite{Bellman:1965}.  That is, $A(s)$ becomes a sequence of step functions where it takes the values $0$ or $1$ and need not satisfy the previous boundary conditions $A(0)=1$ and $A(1)=0$.  This class of solutions to optimal control problems is known as a ``bang-bang'' solution, and is obviously non-adiabatic by construction.  This principle has been shown in quantum optimal control outside of the context of quantum computation, where a Monte Carlo minimization scheme was applied to determine the schedule of step functions, and a different variational principle was employed~\cite{Rahmani:2013}.  However this scheme could be straightforwardly adopted using the variational principle methods described here to engineer state preparation schedules for a state of interest, or to perform more general quantum computation.

\subsection{Unitary coupled cluster}
Another method to parametrically explore the Hilbert space of possible quantum states is the unitary coupled cluster method developed in quantum chemistry~\cite{Taube:2006,Bartlett:2007}.  The projective non-unitary (and non-variational) form of these equations form the basis for the gold-standard of classical quantum chemistry, coupled cluster with single and double excitations with perturbative triple excitations [CCSD(T)]~\cite{Raghavachari:1989,Bartlett:2007} and has its origins in nuclear physics~\cite{Coester:1960}.  The unitary form of these equations do not have a well defined truncation as the projective form does, and one must rely on perturbative arguments to handle the BCH expansion that break down when the parameters defining the states grow.  This ansatz for electronic systems has been documented in classical quantum chemistry and in previous works on the variational quantum eigensolver~\cite{Taube:2006,Bartlett:2007,Peruzzo2014,Yung:2014}, and here we document its generalization to generic collections of interacting two-level quantum systems, which include the anti-symmetric electronic case as a specialization.  We note that coupled cluster has been utilized before in the context of frustrated spin systems such as Kagome lattices~\cite{Schmalfuss:2006,Gotze:2011}, but our treatment will extend beyond a fixed reference and also focus on the unitary variant of the method.

To conceptually introduce the approach, recall the introduction of reference states earlier in this work, and consider a single computational reference state of an $N$-qubit quantum system, $\ket{\Phi_{R0}} = \ket{000...0}$.  One way to parametrically explore Hilbert space is to consider the space of states ``close'' to $\ket{\Phi_{R0}}$ in the sense of Hamming distance or bit flips.  This method, sometimes called configuration interaction (CI) or state space restriction enumerates available states through the use of spin flip~\cite{Helgaker:2014,Kuprov:2007}.  For example, all states $1$ flip away from $\ket{\Phi_{R0}}$ may be written as
\begin{align}
\ket{\Psi_{\text{CI}}(\vec{\theta})} = \sum_{p_1} \theta_{p_1} \sigma^+_{p_1} \ket{\Phi_{R0}}.
\end{align}
where in this case $\theta_{i}$ are complex coefficients and $\sigma_{p}^+$ is the qubit raising operator applied to qubit $p$.  This expansion can be extended systematically by including multi-qubit spin-flip operators to eventually parametrize all states in the Hilbert space, or full configuration interaction (FCI).  While this parametric construction of states is straightforward, it has a number of deficiencies that render it non-optimal.  We will not attempt to explore all of those here, and note only that this ansatz is efficient to prepare and use classically for any truncation to a fixed number of spin flips $k$, and it is not clear that there is an advantage to specifically preparing a linear truncated state on a quantum device.  

An idea closely related to this is coupled cluster, which also uses the spin-flip concept to explore states ``close'' to a reference, but as a generator used in exploration of the space.  In the case of quantum computing, its unitary variant is of particular interest, as unitary state preparation is a natural operation on a quantum computer.  Conventional implementations of coupled cluster often utilize a single, well defined reference state with all spins aligned, i.e. $\ket{\Psi_{R0}} = \ket{000...0}$.  With this assumption, one may explore all of quantum space through successive flips in the computational basis.  As a simple example, if one is interested in only real wavefunctions, the space of single spin flips may be explored by
\begin{align}
\ket{\Psi_{\text{CC1}}(\vec{\theta})} &= \exp \left[ \sum_{p_1} \theta_{p_1} \left( \sigma^+_{p_1} - \sigma^{-}_{p_1}\right) \right] \ket{\Phi_{R0}}
\end{align}
and successively larger fractions of the space of real wavefunctions may be covered by introducing multiple spin flips. In the study of general quantum states however, it is sometimes necessary or more efficient to explore quantum state space from an arbitrary reference $\ket{\Phi_R}$, which could be entangled or simply more complex than $\ket{\Phi_{R0}}$.  These challenges have been studied in the context of multi-reference coupled cluster in quantum chemistry~\cite{Ladig:1984,Musia:2011}.  Moreover in quantum computation one may not have perfect knowledge of the reference state, nor want to require it in their algorithm.  For example the reference state could be prepared by some adiabatic state preparation procedure.  In this situation one could accidentally have as a reference state $\ket{\Phi_R} = \ket{++...+}$ with $\ket{+}=1/\sqrt{2}(\ket{0} + \ket{1}$, from which no state exploration is possible with the above cluster operator.  The space of non-trivial single qubit operators is spanned by $\sigma^+, \sigma^-, \sigma^z, I$.  As such we want to generalize to a set of anti-Hermitian operators spanning the same space, given by
\begin{align}
  i(\sigma_p^+ + \sigma_p^-) = i \sigma^1_p &= \left(\begin{array}{cc}0 & i \\ i & 0 \end{array}\right)_p \\
  (\sigma_p^+ - \sigma_p^-) = i \sigma^2_p &= \left(\begin{array}{cc}0 & 1 \\ -1 & 0 \end{array}\right)_p  \\
  i\sigma^3_p &= \left(\begin{array}{cc}i & 0 \\ 0 & -i \end{array}\right)_p
\end{align}
For convenience we have introduced the standard Pauli operators in the numerical indexing scheme, that is $\sigma^0=I$, $\sigma^1=\sigma^x=X$, $\sigma^2=\sigma^y=Y$, $\sigma^3=\sigma^z=Z$.  As one is not typically interested in global phase factors, we implicitly ignore the identity operator in all equations going forward and with the remaining operators we may write the first order cluster operator as
\begin{align}
T_1(\vec{\theta}) = i \sum_{p_1 \alpha_1} \theta_{p_1}^{\alpha_1} \sigma_{p_1}^{\alpha_1}
\end{align}
where $\theta_{p_j}^{\alpha}$ are real, Roman indices $p_j$ indicate different qubits, and the Greek indices indicate different Pauli operator bases. More generally the $k$'th order cluster operator may be written as
\begin{align}
T_k(\vec{\theta}) = i\sum_{\vec{p}, \vec{\alpha}} \theta_{\vec{i}}^{\vec{\alpha}} \sigma_{\vec{i}}^{\vec{\alpha}}
\end{align}
where $\sigma_{\vec{p}}^{\vec{\alpha}} = \sigma_{p_1}^{\alpha_1}\sigma_{p_2}^{\alpha_2}...\sigma_{p_k}^{\alpha_k}$,  $\theta_{\vec{p}}^{\vec{\alpha}}$ is a $k-$index tensor containing the variational parameters, and the full cluster operator up to order $k$ is written
\begin{align}
T^{(k)}(\vec{\theta}) = \sum_{i}^{k} T_i(\vec{\theta})
\end{align}
From this general cluster operator, we define the unitary coupled cluster state of order $k$ with reference $\ket{\Phi_R}$ as
\begin{align}
\ket{\Psi^{(k)}_{CC}(\vec{\theta})} = \exp \left(T^{(k)}(\vec{\theta})\right) \ket{\Phi_R} 
\end{align}
With this exposition it becomes clear that unitary coupled cluster generators for a totally general spin reference case at order $k$ are the anti-Hermitian algebra $\mathfrak{su}(2^k)$ and the set of possible actions on the qubits are all possible unitary transformations on $k$ qubits that leave the global phase unchanged, or $\text{SU}(2^k)$.

This represents a parametric state preparation with $O((3N)^k)$ real parameters. While this has the potential to represent any known quantum operation at sufficient order and precision of implementation, practically speaking one often restricts to the case of $k=2$, which has been found to be quite powerful in expressing states in quantum chemistry.  This represents a powerful ansatz with a number of parameters that grows only quadratically in the size of the system.  Additionally, the state preparation is manifestly unitary by construction, and has no known efficient classical preparation or method for sampling with arbitrary (possibly entangled) reference $\ket{\Phi_R}$.  As has been noted previously, this state can be prepared efficiently for any fixed order $k$ to a specified accuracy on a quantum device by using the Suzuki-Trotter factorization of the unitary operator $\exp(T^{(k)}(\vec{\theta}))$~\cite{Peruzzo2014,Trotter:1959,Suzuki:1993}.  We note that as one is not trying to faithfully reproduce some dynamics as in many uses of the Suzuki-Trotter factorization, that a coarse factorization may suffice, simply altering the definition of the ansatz, but still remaining difficult to simulate classically.

As an extension to the suggested implementation of spin unitary coupled cluster by Suzuki-Trotter, one may use the connection to $\mathfrak{su}(2^k)$ to take a more geometric approach and explore states through geodesic constructions as was done by Nielsen et al.~\cite{Nielsen:2006}.  Moreover if one allows values of different parameters at different Trotter steps, one may perform arbitrary 1 and 2 qubit gates at $k=2$, which forms a universal gate set and the ansatz can be made equivalent to an arbitrary quantum circuit with a sufficient number of Trotter steps. To see this, consider the first order in a Trotter factorization with a second order cluster operator and a Trotter number of $N$. One could prepare the desired state from a given reference $\ket{\Phi_{\text{ref}}}$ as
\begin{align}
\ket{\Psi_{cc}(\vec{\theta})} = \left[ \prod_{p_1 p_2 \alpha_1 \alpha_2} \exp \left( i \frac{\theta_{p_1 p_2}^{\alpha_1 \alpha_2}}{N} \sigma_{p_1 p_2}^{\alpha_1 \alpha_2} \right) \right]^N \ket{\Phi_{\text{ref}}}
\end{align} 
where we emphasize that it is more correct to consider the use of the exponential splitting as a redefinition of the ansatz than an approximation.  Instead of following this precise splitting procedure, where the same parameters are used in each Trotter step, one can relax the parameters to have independent values at each time step, and to not split Pauli operators acting on the same two qubits within one time step.  This results in an ansatz of the form
\begin{align}
\ket{\Psi_{cc}(\vec{\theta})} = \prod_t^N \left[ \prod_{p_1 p_2} \exp \left( i \sum_{\alpha_1 \alpha_2} \theta_{p_1 p_2}^{\alpha_1 \alpha_2}(t) \ \sigma_{p_1 p_2}^{\alpha_1 \alpha_2} \right) \right] \ket{\Phi_{\text{ref}}}.
\end{align} 
The operator defined by
\begin{align}
O  = i \sum_{\alpha_1 \alpha_2} \theta_{p_1 p_2}^{\alpha_1 \alpha_2}(t) \ \sigma_{p_1 p_2}^{\alpha_1 \alpha_2}
\end{align}
can express an arbitrary element in $\mathfrak{su}(4)$ and thus its exponential $\exp(O)$ can be used to form an arbitrary two qubit gate on any two qubits, or said differently, an arbitrary element of $SU(4)$ on any two qubits.  Arbitrary two qubit gates on any qubit are known to constitute a universal gate set~\cite{Nielsen:2010}, and then clearly can be used to construct any desired universal gate set such as the Clifford$+$T set.  This establishes a clear connection between second order unitary coupled cluster and universal quantum computation through relaxation of parameters in an exponential operator splitting. This also opens the research direction of connecting states of this type to tensor networks where the network is defined by the action at each ``timestep'' of unitary coupled cluster~\cite{Schwarz:2012}.

\subsection{Fermionic UCC}
Due to particular interest in the quantum chemistry and other fermionic problems, it is worth discussing the specialization of this method to those cases.  First taking again the case of a fixed computational reference, such as $\ket{\Phi_{R0}} = \prod_i a_i^\dagger \ket{}$, in analogy to the spin case, the first and second order cluster operators conventionally take on a simple form, that is
\begin{align}
T^{(1)}(\vec{\theta}) &= \sum_{i_1 p_1} \theta_{i_1 p_1} (a_{i_1}^\dagger a_{p_1} - a_{p_2}^\dagger a_{i_1}) \\
T^{(2)}(\vec{\theta}) &= \sum_{i_1 i_2 p_1 p_2} \theta_{i_1 i_2 p_1 p_2} (a_{i_1}^\dagger a_{p_1} a_{i_2}^\dagger a_{p_2} - a_{p_2}^\dagger a_{i_2} a_{p_1}^\dagger a_{i_1}) 
\end{align}
with $i_j$ indexing the occupied orbitals, $p_j$ indexing the unoccupied orbitals, and higher orders defined in the obvious way of including more excitation operators.  These generators are constructed to conserve particle number at all orders and parametrically depend on $O(M^{2k})$ real parameters at order $k$. In the case of a single reference, it should be noted that all the excitation operators commute as a direct consequence of the creation and annihilation operators being restricted to act on different subspaces.  As a result, Trotter factorization of this ansatz may be performed to arbitrary times exactly that allows one to explore regimes where low order truncations of the BCH expansion are not accurate and thus may be difficult to sample from classically.

We can understand the equivalent action on qubits by mapping the fermion operators to spin operators via either the Jordan-Wigner or Bravyi-Kitaev transformations discussed earlier in this work.  In the case of the Jordan-Wigner mapping, as a result of the non-locality of these mappings, at every fermion order $k$, we find spin flips up to all $N$ spins and observe that the allowed operations on the qubits are a non-trivial subgroup of $SU(2^k)$ at every order $k$.  This demonstrates that it is key to develop the ansatz in the fermionic framework before mapping the problem to a spin representation.  If one were to first map to spins, then use the spin coupled cluster formulation, the ansatz might explore many irrelevant or symmetry broken states, such as mixtures of different particle number states.

In analogy to our exposition on spins however, this type of cluster operator is reference state specific.  That is, there are some reference states from which it will fail to parameterize the entirety of the $N$ fermion space and extensions to multi-reference states can require a different cluster operator for each reference.   This can be seen from dimension counting in the vector space of the fermion excitation operators.  For example at first order these operators only span a real vector space of dimension $M^2/2 - M$ whereas the full space of all $1$ fermion linear operators has real dimension $M^2$.  In classical implementations of multi-reference coupled cluster there are many different approaches to solving this and related problems going by names such as ``universal'' or ``state selective'' multi-reference coupled cluster~\cite{Ladig:1984,Musia:2011}.  In the case of unitary coupled cluster on a quantum computer, in analogy to how we generalized the distinguishable spin operators, we can generalize the fermion operators to treat arbitrary references without such concerns.

The operators $a_i^\dagger a_j$ and their tensor products where $i$ and $j$ run over all $M$ orbitals (instead of restricting them to occupied and unoccupied relative to a reference) form a basis for the real vector space of operators on $N$ fermion states. As a result, to allow arbitrary action on the space of $N$ fermions, the span of the generating operators used must match this.  To span the same real vector space as these operators we use the following anti-Hermitian basis
\begin{align}
  i (a_p^\dagger a_q + a_q^\dagger a_p) = i A_{pq}^1 \ & ; 1 \leq p \leq q \leq M \\
  a_p^\dagger a_q - a_q^\dagger a_p = i A_{pq}^2 \ & ; \ 1 \leq p < q \leq M
\end{align}
and all possible $N-$fold tensor products of these operators.  One can verify by dimension counting of the real vector space that these operators in fact span the entire space of possible fermion operators.  With these operators, the first order fermion cluster operator can be written as
\begin{align}
  T_1(\vec{\theta}) = i \sum_{p_1 q_1\alpha} \theta_{p_1 q_1}^{\alpha} A_{p_1q_1}^\alpha
\end{align}
where $p_j$ and $q_j$ run over all orbitals and $\alpha$ indexes the anti-Hermitian fermion generators.  Higher orders of the cluster operator can be built naturally from tensor products of these operators, such that at the $k$'th order we have
\begin{align}
  T_k(\vec{\theta}) = i\sum_{\vec{p},\vec{q},\vec{\alpha}} \theta_{\vec{p} \vec{q}}^{\vec{\alpha}} A_{\vec{p} \vec{q}}^{\vec{\alpha}}
\end{align}
where the same vector operator shorthand as the spin case has been used.  With this construction the power of the cluster operator is state agnostic, and fermion number conserving.  We term this the state universal quantum unitary coupled ansatz (SU-QUCA). Again, in all cases the optimal choice of the parameters $\vec{\theta}$ is determined through the application of the variational principle with respect to the Hamiltonian of interest.

\subsection{Quantum Error Suppression and Symmetries}
A variational hybrid quantum-classical is designed to perform on pre-threshold computers, where gates may be imperfect and random bit flip or phase errors may be introduced into the computation. Fortunately the variational formulation allows one to suppress certain types of errors naturally, which we will discuss here in the context of variational error suppression.

In the design of a parametric wavefunction ansatz, it is common to enforce known symmetry requirements for both theoretical and practical purposes.  For example, in the fermionic unitary coupled cluster wavefunctions, the ansatz is designed to conserve the number of particles for all possible choices of the parameters $\vec{\theta}$. That is both the ansatz and the Hamiltonian commute with the number operator $N=\sum_i a_i^\dagger a_i$. While we haven't explicitly done so here, it is also possible to adapt the cluster operators to conserve total spin~\cite{Helgaker:2014}. In a fully error corrected quantum computer, this introduces no additional concerns and can simplify the problem under consideration. However in a pre-threshold device or any with only partial error correction this must be taken into consideration.

Consider the preparation of an ansatz from some initial state, which we denote as $U_a(\vec{\theta})$.  In a pre-threshold, non-error corrected quantum device, there can be a distinction between the formal specification of the ansatz preparation $U_a(\vec{\theta})$ as a gate or operation sequence and the operation sequence actually performed on the system with inputs $\vec{\theta}$, which we will denote $\tilde{U}_a(\vec{\theta})$.  We call an error in such an implementation suppressible if there exists a correction input vector $\vec{\beta}$ such that $||U_a(\vec{\theta}) - \tilde{U}_a(\vec{\theta}+\vec{\beta})|| < \epsilon$ for a specified $\epsilon > 0$, and further denote it variationally suppressible if the corrected vector $\vec{\theta} + \vec{\beta}$ also corresponds to an optimum on the parameter surface.  In such a case, the variational quantum eigensolver can suppress these errors naturally without detailed knowledge of the error mechanism.
\begin{figure}[t!]
  \centering
    \includegraphics[width=8cm]{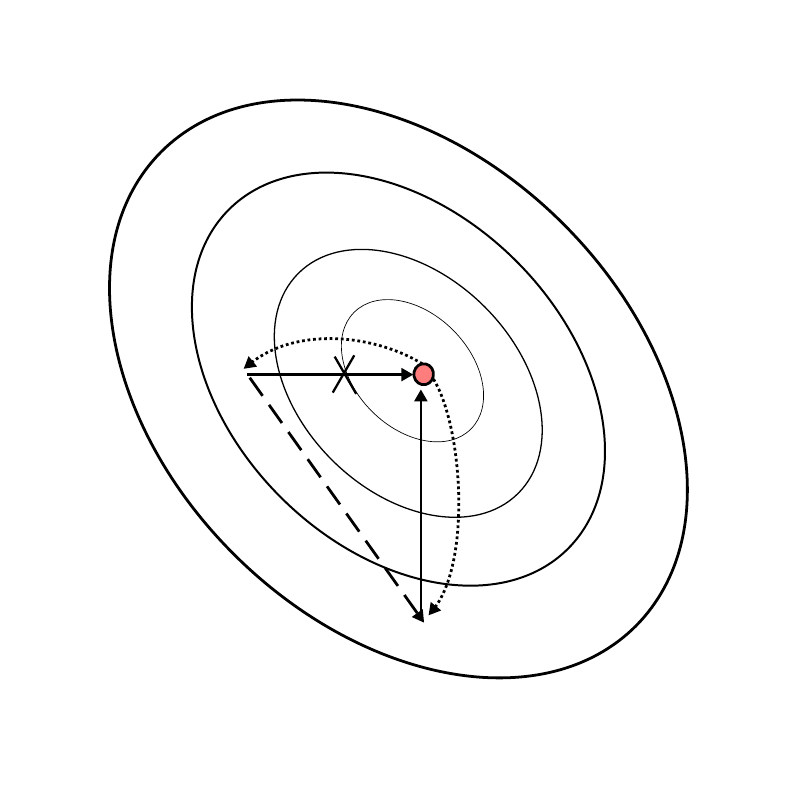}
      \caption{A cartoon depicting the concept of variationally suppressible errors on energy contours.  Dotted lines represent errors that move the state away from the variational minimum, and solid lines characterize a shift of the ansatz parameters that can return the state to the minimum.  In this case the vertical axis is within the manifold of the ansatz parameters, while the horizontal axis is not, as indicated by the cross in the line returning along that axis.  However by adding additional operators, represented by the diagonal dashed line, it becomes possible to suppress these errors variationally.} 
\end{figure}
A troublesome non-suppressible case is when an error violates a symmetry of the ansatz.  More explicitly, if we denote the symmetries of the ansatz as the set of operators $S$ such that $[U_{a}(\vec{\theta}), S]=0$ for all $\vec{\theta}$, then for any symmetry violating error $U_e$ such that $[U_e, S] \neq 0$, there does not exist any correction vector $\vec{\alpha}$ such that the desired preparation can be performed.  

To be more concrete, consider the two examples given in this section, parameterized adiabatic state preparation and coupled cluster.  In these cases, some symmetries of the ansatz can be trivially determined by the generating operators. In adiabatic state preparation, the symmetries will be given by the set of operators $S$ such that $[H_i, S]= 0$ for all Hamiltonians $H_i$, including the initial, problem, and intermediate Hamiltonians.  In the case of coupled cluster, this will be the set of operators $S$ such that $[E_i, S]=0$ for all excitation type operators $E_i$, such as the number operator.  These represent sufficient conditions for $[S, U_a(\vec{\theta})]=0$ for every possible choice of $\vec{\theta}$.  In the case of fermionic coupled cluster, the generating operators are specifically designed to conserve particle number, such that one symmetry of the system is the number operator $N=\sum_i a_i^\dagger a_i$.  In a Jordan-Wigner qubit representation, this simply counts the number of qubits in state $\ket{0}$.  As such, if a random error of the form $U_e = c_1 \sigma^1$ is acted on any qubit, this error is not suppressible.

This particular error can be made suppressible by extending the set of generating operators to include spin flips (e.g. $i \sigma^+_p$ and $i \sigma^-_p$) or fermionic non-number conserving operators, e.g. $(a_p^\dagger - a_q)$ and $i (a_p^\dagger + a_q)$ as well as all tensor products of these operators with the rest of the generating set.  With the addition of these operators, this error become suppressible, however the error will only be variationally suppressible if the desired symmetry state of the ansatz corresponds to an energetic minimum.  In the event that it does not, one can construct an auxiliary Lagrangian of the form
\begin{align}
\mathcal{L} = H + \sum_i \lambda_i (S_i - s_i I)^2
\end{align}
where $\lambda_i$ are penalty multipliers and $s_i$ are constants corresponding to the desired expectation values of the operators $S_i$.  In order to be efficient, measurements corresponding $S_i^2$ and $S_i$ must be also be efficient.  Using this construction, one may minimize with respect to expectation values $\avg{\mathcal{L}}(\vec{\theta})$ instead of $\avg{H}(\vec{\theta})$, and in the limit that $\lambda_i \rightarrow \infty$ the symmetries will be exactly preserved while allowing variational error suppression under action by the extended operator set.  

This methodology also allows for access to excited states that correspond to an energetic minima of a given symmetry.  An example of this could be the lowest triplet energy state of a molecule with a natural singlet ground state, or the ionic state of a molecule after photodissociation.  Use of this construction may allow easier access to these particularly important excited states, as compared to a more general excited state approach.

\section{Operator Averaging}
Once a trial state $\ket{\Psi(\vec{\theta})}$ has been prepared, the next crucial step in the VQE is the evaluation of the objective function corresponding to the problem operator $H$, $\avg{H}(\theta) = \bra{\Psi(\vec{\theta})} H \ket{\Psi(\vec{\theta})}$.  One possibility is to use the quantum phase estimation algorithm~\cite{Abrams1997,Abrams1999,Aspuru:2005}.  If $\ket{\Psi\vec(\theta)}$ is an eigenstate, then the value is obtained after a single state preparation with a cost in the desired precision of $O(1/\epsilon)$.  Unfortunately, to achieve this precision, all of the operations must be coherent which is a prohibitive technological requirement for current and near-term quantum computers.  Moreover, if the state is instead a mixture of many eigenstates, it will still require $O(1/\epsilon^2)$ repetitions of the entire procedure to converge the value $\avg{H}(\theta)$ to a precision $\epsilon$.  The use of quantum phase estimation done to a precision surpassing $\epsilon$ opens the possibility to instead minimize the minimal value found in a projective measurement of the energy in a sequence of phase estimation runs.  However we do not explore that option further here.

In 2014, Peruzzo and McClean et. al ~\cite{Peruzzo2014} suggested a way to retain the advantage of preparing classically inaccessible states while removing the overwhelming coherence time requirements to measure the energy.  This method is called Hamiltonian averaging and has been discussed recently in more detail~\cite{McClean:2014}.

The original formulation used the fact that tensor products of Pauli operators form a basis for the space of Hermitian operators.  As such any Hermitian operator $H$ may be written as
\begin{align}
H = \sum_{i_1\alpha_1} h^{i_1}_{\alpha_1} \sigma^{i_1}_{\alpha_1} + \sum_{i_1 i_2 \alpha_2 \alpha_2} h^{i_1 i_2}_{\alpha_1 \alpha_2} \sigma^{i_1 i_2}_{\alpha_1 \alpha_2} + ...
\end{align}
and by linearity the expectation value as
\begin{align}
\avg{H}_{\ket{\Psi}} = \sum_{i_1\alpha_1} h^{i_1}_{\alpha_1} \avg{\sigma^{i_1}_{\alpha_1}} + \sum_{i_1 i_2 \alpha_2 \alpha_2} h^{i_1 i_2}_{\alpha_1 \alpha_2} \avg{\sigma^{i_1 i_2}_{\alpha_1 \alpha_2}} + ...
\end{align}
As a result, all that is required is the weighted sum of the results from simple Pauli measurements.  This is an operation requiring coherence time $O(1)$ assuming parallel qubit rotation and readout are possible, otherwise the coherence time required is $O(k)$ where $k$ is the locality of the term to be measured. Previously, some scaling analysis of this procedure was done in the context of locality~\cite{McClean:2014}, but here we detail more specifically how to perform the averaging and verify the error on the fly in a simulation of a general state.

Consider the Hamiltonian decomposed as
\begin{align}
H = \sum_{\gamma} H_{\gamma}.
\end{align}
where each $H_\gamma$ is a Hermitian operator with associated measurement outcomes $m_1$ and $m_2$, of which Pauli operators are a special case.  In order to get the desired precision in a normal distribution approximation, we require a variance of $\epsilon^2$ in the estimator of $\avg{H}$, which we denote with a large hat as $\widehat{\avg{H}}$.  The estimator we have described is constructed as a sum of independent estimators $\widehat{\avg{H_\gamma}}$, 
\begin{align}
\widehat{\avg{H}} = \sum_{\gamma} \widehat{\avg{H_{\gamma}}}
\end{align}
each of which is a built a sequence of independent measurements $X$.  As the measurements are taken from independent state preparations, we have that the covariance between the individual estimators on the measurements is $0$ or $\text{Cov}[\widehat{\avg{H_\alpha}}, \widehat{\avg{H_\beta}}] = 0 \ \forall \ \alpha \neq \beta$ and thus the variance of the total estimator is the sum of the variances of the individual estimators
\begin{align}
\Var{\widehat{\avg{H}}} = \sum_\gamma \Var [\widehat{\avg{H_\gamma}}].
\end{align}
The individual estimators are constructed as the mean of a sequence of independent measurements corresponding to the operator $H_\gamma$ on independent preparations of the state $\rho$.  Each measurement of the total operator requires a state preparation and measurement for each individual term, and thus the total number of expected state preparations and measurements to achieve a precision of $\epsilon$ in $\widehat{\avg{H}}$ is
\begin{align}
n_{\text{expect}} = M \sum_{\gamma} \frac{\Var[H_\gamma]}{\epsilon^2}.
\end{align} 
While this offers insight into how many measurements one expects to take, it does not yet constitute a practical algorithm, as the true value of the variances $\Var[H_\gamma]$ in general will be unknown except in toy examples.  Instead, one has access to the sample mean and unbiased sample variance as the measurements are taken.  That is, after $n$ measurements $\{ x_i \} $ of the operator $H_\gamma$ have been taken on $\rho$, one computes
\begin{align}
\widehat{\avg{H_{\gamma}}} ( \{ x_i \} ) &= \frac{1}{n} \sum_i^n x_i \notag \\
\widehat{\Var[H_{\gamma}]} ( \{ x_i \} ) &= \frac{1}{n-1} \sum_i^n(x_i - \widehat{\avg{H_{\gamma}}} ( \{ x_i \} ))^2
\end{align}
and continues taking measurements until $\Var[\widehat{\avg{H_\gamma}]} \approx \widehat{\Var[H_{\gamma}]} ( \{ x_i \} )/n < \epsilon^2/M$, and moves on to the next term.  While straightforward, this methodology suffers from some ambiguities when using a small number of measurements or when the state $\rho$ represents an eigenstate of the operator $H_\gamma$.  In particular, how many measurements are required to confirm that the variance is $0$ to the desired precision.  This is related to how unobserved events are addressed in a frequentist perspective of probability.  In practical implementations these issues are often left unaddressed rigorously in stochastic sampling methods and a reasonable minimum number of measurements is chosen such as $n=1000$ or $n=10000$ before the estimates of $\widehat{\Var[H_{\gamma}]} ( \{ x_i \} )$ are taken to be reliable, trusting that after a number of samples that it is well represented by a normal distribution and the higher moments associated with errors in estimates of the variance vanish rapidly.  An alternative perspective that addresses such concerns from the outset is a Bayesian perspective, which has been investigated in the context of quantum phase estimation~\cite{Wiebe:2015}, and we now explore in the context of Hamiltonian averaging.

\subsection{Bayesian Perspective}
In a Bayesian perspective, we start from an uninformative prior for the distribution $\widehat{\avg{H_{\gamma}}}$.  In the case of two measurement outcomes, the likelihood function is the binomial likelihood, and the posterior distributions after measurement can be worked out analytically when used with a conjugate Beta prior.  These distributions are well-defined even for small numbers of measurements or when $\rho$ is close to an eigenstate of $H_\gamma$, resulting in potentially unobserved events in a sequence of measurements.  

Consider a sequence of independent measurements $X$ with two possible outcomes $\{m_1,m_2\}$, such as the quantum measurement of a Pauli operator.  The likelihood of observing the sequence of measurements $X$ is completely defined by a single variable $p$, and is written
\begin{align}
P(X|p) = \left( \begin{array}{c} N \\ r \end{array} \right) p^r(1-p)^{N-r}
\end{align}
with $N$ being the total number of measurements $X$ and $r$ being the number of measurements equal to $m_1$.  The value $p$ defines the probability of observing $m_1$ and will be directly related to $\widehat{\avg{H_\gamma}}$. Our current knowledge of $p$ is defined by the prior distribution $P(p)$.  Many choices for the form of the prior distribution can be made, but an analytical result can be obtained by choosing the conjugate prior to the Binomial distribution, which is the Beta distribution
\begin{align}
P(p; \alpha, \beta) = \text{Beta}(\alpha, \beta) = \frac{\Gamma(\alpha + \beta)}{\Gamma(\alpha) \Gamma(\beta)} p^{\alpha-1} (1-p)^{\beta - 1}.
\end{align}
The Beta distribution is a function of two parameters $\alpha$ and $\beta$, and these are the parameters we will seek to update with a Bayes inference scheme.  Simply put, given the measurements $X$ with $r$ instances of $m_1$, the posterior distribution is given by
\begin{align}
\label{eq:posterior}
P(p| X) = \text{Beta}(\alpha+r, \beta+N-r) = \text{Beta}(\alpha', \beta')
\end{align}
From $\alpha'$ and $\beta'$, one can determine both the mean value and variance in our desired quantity as
\begin{align}
\avg{p} &= \frac{\alpha}{\alpha + \beta} \\
\text{Var} [p] &= \frac{\alpha \beta}{(\alpha + \beta)^2(\alpha + \beta + 1)}
\end{align}
and the expected value and variance of $p$ may be used in the estimators associated with $H_\gamma$.  In particular
\begin{align}
\widehat{\avg{H_\gamma}} &= \avg{p} m_1 + (1-\avg{p})m_2 \\
\Var[\widehat{\avg{H_\gamma}]} &= (m_1 - m_2)^2 \Var[p]
\end{align}
A reasonable choice of initial prior in this situation before any measurements are taken is the uniform prior (sometimes called the Bayes' prior probability in this case) $\text{Beta}(1,1)$.  Thus a practical strategy in the Bayes setting is to let $\alpha=\beta=1$, then take $N$ measurements.  One then updates $\alpha$ and $\beta$ to $\alpha'$ and $\beta'$ according to eq. \ref{eq:posterior}, and continues taking measurements until $\Var[\widehat{\avg{H_\gamma}]} < \epsilon^2/M$, which is simply computed as a function of the new $\alpha$ and $\beta$ through the above formulae.  We note that if one has a good reference state, a prior distribution can be constructed from it to yield an informative prior.  This has the potential to reduce the cost and will converge to the same result under most reasonable conditions.  However one must be careful as this may introduce a bias for poor reference states with a small number of measurements

After using either the frequentist or Bayesian approach to check convergence of  $\Var[\widehat{\avg{H_\gamma}]}$ for all $\gamma$, under a normal distribution approximation the final estimation of $\avg{H}$ is precise to the desired precision $\epsilon$.

An alternative to the normal approximation confidence intervals may be used in the Bayesian approach if desired.  As the measurements are taken for each of the operators $H_\gamma$ in the Bayesian approach, the associated probability distribution $P(\widehat{\avg{H_\gamma}})$ is known.  The probability distribution of a sum of independent random variables is known to be the convolution of the individual probability distributions, such that 
\begin{align}
P(\widehat{\avg{H}}) = \Conv_\gamma P(\widehat{\avg{H_\gamma}})
\end{align}
Unfortunately the convolution of two Beta distributions does not have a known analytical result, and these convolutions must be performed numerically.  Once the probability distribution $P(\widehat{\avg{H}})$ is known, one may numerically bracket the desired confidence interval to determine the precision of the approach.  Practically speaking, the convergence of this final probability distribution to a normal distribution is quite rapid, and thus the normal approximation relying on the variance is the standard procedure.

\subsection{Cost Reduction}
The computational cost of Hamiltonian averaging can be reduced in a number of ways.  In this section we will consider two methods for doing so.  In the first we will remove terms that are deemed unimportant, and in the second we will consider how terms are grouped in order to reduce the required number of state preparations.
\subsubsection{Term Truncation}
The first strategy to reduce the number of measurements and state preparations required is to avoid measurements guaranteed not to contribute at the desired precision to the total estimate.  To do this, one may order the terms by their expected maximum contribution to the estimate.  For example the magnitude of a weighted Pauli operator $H_\gamma = h_\gamma \sigma$ is bounded such that for any state $\rho$, $|\avg{H_\gamma}| \leq |h_\gamma|$.   Once the terms are ordered according the the maximum expected contribution, with the maximum at $\gamma = M$, we can construct the sequence of partial sums
\begin{align}
e_k = \sum_{i}^k |h_i|
\end{align}
with $e_0$ defined to be $0$, that defines the maximal bias introduced by truncating the $k$ smallest terms.  Using this sequence, one may choose a constant $C \in [0,1)$ and remove the $k^*$ lowest terms by finding the maximal index $k^*$ in the sequence such that $e_{k^*} < C \epsilon$.  In this choice, $C$ determines the both the number of terms one is allowed to neglect and amount of bias introduced.  As the estimator is now biased, one must consider the bias-variance tradeoff to maintain the desired accuracy.  In order to achieve an expected mean-square-error of $\epsilon$ in the final answer, we must decrease the variance of the estimator on the remaining terms such that $C^2 \epsilon^2 + \sum_\gamma^{M-k^*}\Var[ \widehat{\avg{H_\gamma}}] < \epsilon^2$.  This may be achieved by changing the per-term variance threshold for each $\widehat{\avg{H_\gamma}}$ to be $(1-C^2)\epsilon^2/(M-k^*)$.  This results in a new expected number of measurements 
\begin{align}
n^*_{\text{expect}} = \sum_{\gamma}^{M-k^*} \frac{(M-k^*) \Var[H_\gamma]}{(1-C^2)\epsilon^2}.
\end{align}
One is free to choose a value of $C \in [0,1)$ to maximize computational efficiency according to the particular constraints of experiment and the distribution of operators in the sum.  It has been seen previously that using this strategy in conjunction with locality information can potentially reduce the costs of quantum chemistry calculations dramatically~\cite{McClean:2014}.

\subsubsection{Commuting Groups and Correlated Sampling}
Another strategy one may use besides truncation is to take advantage of commuting operators within the sum to reduce the number of state preparations required.  If two operators $H_\alpha$ and $H_\beta$ commute, they may be measured in sequence on the same state preparation without biasing the final result of the expectation values.  As the state preparation is expected to be more expensive than projective measurements, this has the potential to offer significant savings.  However, the application of this technique requires some care.

While grouping terms into commuting sets cuts down on the number of state preparations required for a single pass at the measurements and does not bias the expected outcome, there is some detail to consider in the statistics of measurement and estimation of uncertainty.  As terms within a commuting set are measured on the same state within each pass of the procedure, two operators within a set may be correlated such that the estimators of their average may have non-zero covariance i.e. $\text{Cov}[\widehat{\avg{H_\alpha}}, \widehat{\avg{H_\beta}}] \neq 0$.  This additional covariance can either require more measurements for the set of terms if the covariance is positive, or less if it is negative in analogy to the method of antithetic variables or correlated sampling in classical Monte Carlo simulations~\cite{Hammersley:1956,Kalos:2008}.  Thus one must be careful to group only operators that result in a practical efficiency gain.  This concept is best illustrated with a short example.

Consider the $2$ spin Hamiltonian
\begin{align}
H = - (X_1X_2 + Y_1Y_2) + Z_1Z_2 + Z_1 + Z_2
\end{align}
where $X,Y,Z$ are the standard Pauli operators and a quantum state
\begin{align}
\ket{\Psi} = \ket{01}
\end{align}
which we will be measuring.  The operators in this Hamiltonian can be grouped in a number of ways into groups of commuting terms.  Consider the following three options
\begin{align}
1.& \ \{-X_1 X_2\}, \{-Y_1 Y_2\}, \{Z_1 Z_2\}, \{Z_1\}, \{Z_2\} \notag \\
2.& \ \{-X_1 X_2\}, \{-Y_1 Y_2, Z_1 Z_2\}, \{Z_1, Z_2\} \notag \\
3.& \ \{-X_1 X_2, -Y_1, Y_2, Z_1 Z_2\},\{Z_1, Z_2\}. \notag
\end{align}
Using the formulas from the previous section to compute the expected number of state preparations for each grouping of operators to a precision $\epsilon$, we may proceed as follows. The expected estimator variance of the first grouping is 2, but prescribes a total number of state preparations per term to be 5 (from 5 sets of commuting operators), resulting in an expected number of state preparations $n_{\text{expect-1}} = 10/\epsilon^2$. In the second case, we maintain the same variance, but group commuting operators together that have 0 covariance, so the number of preparations per iteration is reduced to $3$ and we find $n_{\text{expect-2}} = 6/\epsilon^2$.  The last case has the smallest number of commuting groups, but introduces an extra covariance term that results from covariance between $X_1 X_2$, and $Y_1Y_2$ on the state $\ket{\Psi}$.  As a result, the total number of expected preparations is given by $n_{\text{expect-3}} = 8/\epsilon^2$.  Thus while the last prescription had the fewest number of commuting terms, the second was a better grouping, reducing cost by almost a factor of 2 from the na\"ive measurement of all terms individually.

This simple example illustrates how savings can be achieved through careful grouping, but also highlights the state and operator dependence of this strategy.  The most crucial piece of information in deciding whether to group commuting terms is the covariance of different operators on the state.  If one has a good approximation of the state, this can be estimated classically before an experiment to group operators that are expected to give cost savings.  Alternatively, if one expects many points in an optimization to be similar, this can be estimated once on the quantum state before beginning to a low precision, and these heuristic groupings can be used for the remainder of the experiment.  Again, we emphasize that this strategy will not bias the final result, even if the sets chosen are non-optimal. It is merely a means of sampling cost reduction.

Regardless of the strategy chosen, it is crucial to correctly determine the statistical uncertainty of the final estimate.  One could estimate the covariances from the measurements and account for this, but a perhaps conceptually simpler approach more true to the spirit of the experiments is to define new trivial estimators $\widehat{\avg{Q_i}}$, which are constructed as follows.  After a state preparation, each operator in $Q_i$ is measured in turn in some pre-defined order to give a sequence $\{x^\gamma_i\}$.  The sum of these measurements for all the operators is defined to be the new measurement $q_i = \sum_\gamma x_i^\gamma$, and the estimator for the average over many realizations is simply the arithmetic mean, $\widehat{\avg{Q_i}} = \frac{1}{n}\sum_j^n q_j$.  In this way the final estimator may be constructed equivalently as
\begin{align}
\widehat{\avg{H}} = \sum_i \widehat{\avg{Q_i}}
\end{align}
that clearly yields the same expectation value but is now composed of estimators such that $\text{Cov}[\widehat{\avg{Q_i}}, \widehat{\avg{Q_j}}] = 0$ for $i \neq j$, allowing one to more conveniently estimate only variance of uncorrelated estimators to determine the uncertainty in the final estimate and fix the desired tolerances per term when measuring.

\subsection{Beyond Energy to General Observables}
Finally we note that the method of calculating operator averages outlined in this section often yields additional information besides the original designed expectation value.  For example, in the case of quantum chemistry, the individual operators measured that compose the Hamiltonian are the reduced $1$ and $2$ electron density matrices, defined for a state $\ket{\Psi}$ as
\begin{align}
D^{i}_{p} &= \bra{\Psi} a_i^\dagger a_p \ket{\Psi} \\
D^{ij}_{pq} &= \bra{\Psi} a_i^\dagger a_j^\dagger a_p  a_q \ket{\Psi}.
\end{align}
Knowledge of these reduced density matrices is sufficient to determine not only the energy but the expectation value of any one- and two-electron operators, such as the dipole moment or charge density.  This follows from the fact that any one- and two-electron operators $F$ and $G$ may be written in a basis as
\begin{align}
F &= \sum_{ip} f_{ip} a_i^\dagger a_p \\
G &= \sum_{ijpq} g_{ijpq} a_i^\dagger a_j^\dagger a_p a_q
\end{align}
where $f_{ij}$ and $g_{ijkl}$ are precomputed with the single particle basis set.  From this it is clear that the expectation values are
\begin{align}
\avg{F} &= \sum_{ip} f_{ip} \bra{\Psi} a_i^\dagger a_p \ket{\Psi} 
        = \sum_{ip} f_{ip} D^{i}_{p} \\
\avg{G} &= \sum_{ijpq} g_{ijpq} \bra{\Psi} a_i^\dagger a_j^\dagger a_p a_q \ket{\Psi} 
        = \sum_{ijpq} g_{ijpq} D^{ij}_{pq}
\end{align}
which may be computed trivially on a classical computer with the measured values from experiment.  Thus the operator averaging methodology in this section gives access to a number of interesting observables of the quantum system with no additional required measurements, and this approach can be viewed alternatively as a form of scalable partial tomography.  This point of view also suggests that a promising route for additional post-processing of data is to use techniques designed to enforce physical constraints on the estimated reduced density matrices~\cite{Banaszek:1999}.  

\section{Optimization of $\vec{\theta}$}
The final piece of the variational quantum eigensolver is a method for updating the parameters $\vec{\theta}$ based on the measured value of the objective function of interest.  The dependence of the objective function on the parameters will, of course, depend upon the ansatz being used and will in general be non-linear and non-convex.  This is not to say ansatz satisfying desirable criteria such as convexity could not be designed, but rather that in general it may not be.  As such, one may not expect global optimization or verification of a proposed solution to be feasible.  However, in many cases local optima are sufficient and prior knowledge of a problem offers high quality starting points for the optimization.  This has often been the case in quantum chemistry, where non-linear procedures such as Hartree-Fock utilize very good local optima and benefit greatly from high quality starting guesses.  The use of high quality starting guesses will likely be important for all types of ansatz discussed here as well.  In the case of UCC for example, perturbation theory methods such as MP2 could be used to generate starting guesses.

\begin{figure}[t!]
  \centering
    \includegraphics[width=8cm]{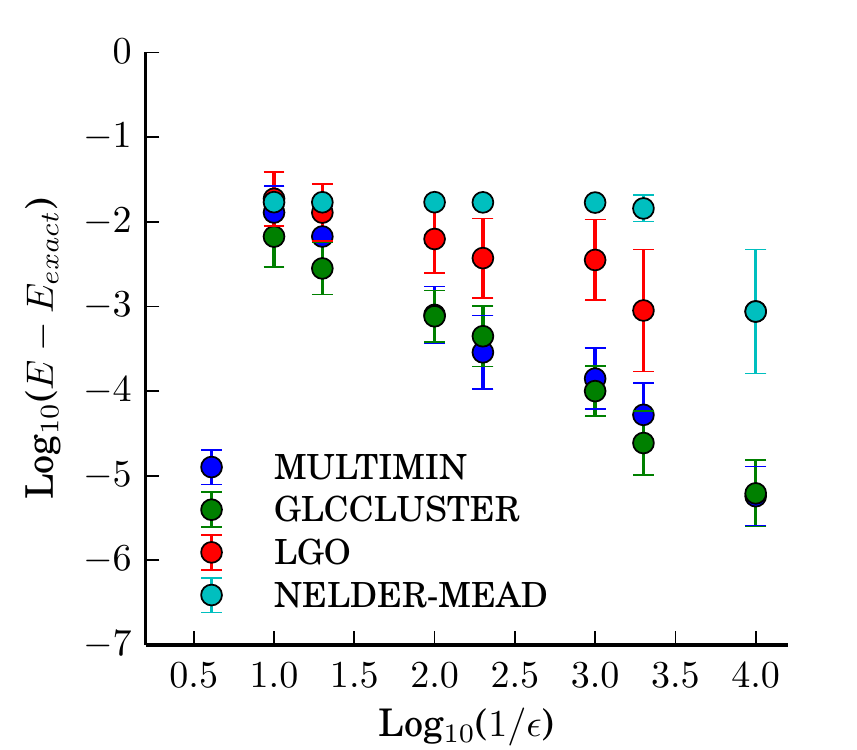}
    \caption{The accuracy of the final energy of the optimized wavefunction at convergence compared to the known exact solution, as a function of the precision in the function value in the optimizer for different methods ($\epsilon$).  The values are averaged over 20 repetitions and the error bars indicate 1 standard deviation of the measured data.  The TOMLAB methods provide dramatically superior performance at essentially all levels of measurement precision above $\epsilon=.1$. \label{fig:Accuracy}}
\end{figure}

The field of non-linear optimization is well developed with many tools both general and more specialized methods to different optimization problems~\cite{Nocedal:2006}.  The objective function by design here is statistical in nature, making it difficult to directly use many of the basic tools from numerical optimization that rely on gradients.  In the original implementation, the derivative free Nelder-Mead simplex method was used as it has reasonable robustness to small quantities of noise, at least in comparison to methods such as standard gradient descent.  However, with developments in the optimization of functions, it is clear that there are more efficient options available for this problem and in this work we compare the Nelder-Mead simplex method, TOMLAB/GLCLUSTER, TOMLAB/LGO, and TOMLAB/MULTIMIN methods~\cite{Rios:2013,Holmstrom:2015} for an example problem.  These particular algorithms were chosen because of Nelder-Mead's use in the original work, and the superior performance of the TOMLAB algorithms in a recent comprehensive benchmark of derivative free optimization techniques~\cite{Rios:2013}.  Each of the TOMLAB algorithms uses a different derivative free search strategy and include both global and local considerations in the choice of new iterates.  Details of the TOMLAB algorithms can be found in the user's guide~\cite{Holmstrom:2015}.

\begin{figure}[t!]
  \centering
    \includegraphics[width=8cm]{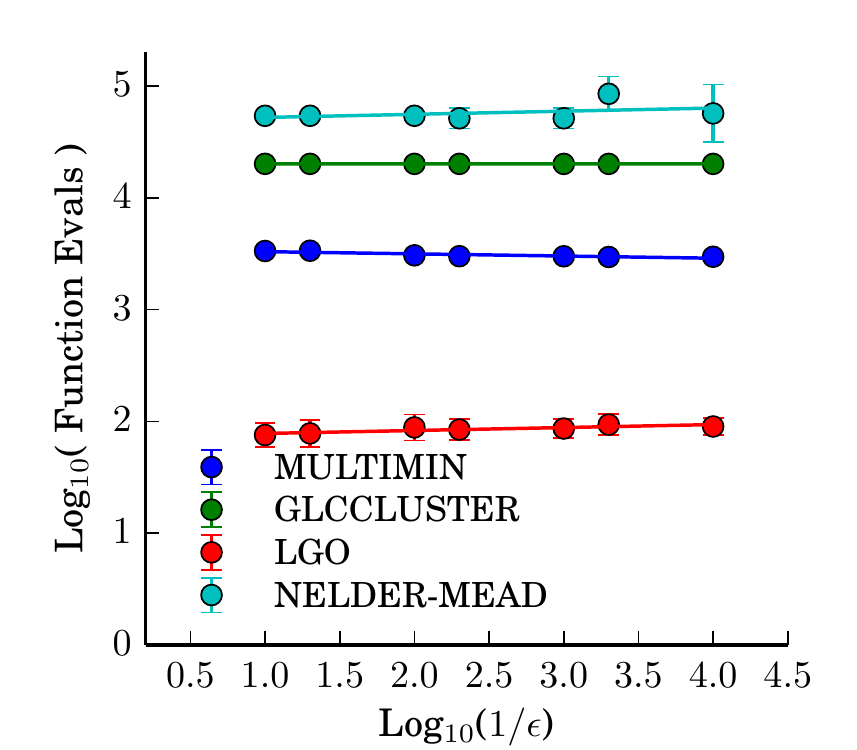}
    \caption{The number of function evaluations required to reach convergence for minimization of the wave function as a function of the precision in the function value.  The accuracy of each of these minimizations relative to the exact answer is shown in Fig. \ref{fig:Accuracy}.  The TOMLAB methods are seen to be dramatically more efficient than the Nelder-Mead method, requiring sometimes 3 orders of magnitude less function evaluations to achieve higher accuracy in the final answer for higher desired precisions.   \label{fig:FEVals}}
\end{figure}

The example problem we benchmark is this case is the optimization of a unitary coupled cluster wavefunction for H$_2$ in a minimal STO-3G basis.  In these benchmarks, simulated measurement estimator noise is added to the objective function at a specified variance $\epsilon^2$.  The optimization is then repeated $20$ times at a given $\epsilon$ and the resulting accuracy with respect to the exact solution is plotted in Fig \ref{fig:Accuracy} as a function of the measurement noise, which can be controlled through the number of measurements taken in the experiment.  The error bars indicate 1 standard deviation in the distribution of values measured over the 20 repetitions.  Additionally, the number of evaluations of the expectation value of the energy required to reach convergence is plotted as a function of the same precision $\epsilon$. It is seen in these plots that in all instances, the TOMLAB methods not only converge to a higher accuracy in the energy, but do sometime as many as 1000 times less function evaluations than the Nelder-Mead method which was previously coupled to the variational hybrid quantum-classical approach.  Moreover, the approximately constant number of function evaluations required to reach convergence as a function of precision suggests that more savings may be reached by using a variable precision optimization, as the cost of a function evaluation to a precision $\epsilon$ scales roughly as $1/\epsilon^2$ in this case.  

While the performance of the TOMLAB algorithms is impressive relative to previous standards, these methods that utilize some global optimization and random search strategies will require further numerical testing as the dimension of the problem space grows.  Moreover, none of these methods were specifically designed for a stochastic objective function.  This is an area of great importance in the algorithm as a whole, and all improvements can translate to dramatic savings in the overall runtime.  As a result this is a topic of ongoing research.

\section{Conclusions}
Quantum computers promise to change the way we think about problems across a plethora of different fields, including the important areas of optimization and eigenvalue problems.  While the construction of full scale, error corrected quantum devices still poses many technical challenges, great progress is being made in their development.  In the era of pre-threshold devices, and indeed beyond it, quantum devices may find an advantage in leveraging classical resources alongside quantum resources to exploit the powerful technologies already in existence today. The variational quantum eigensolver is an algorithm designed to exploit these resources in both a pre- and post-threshold world, and it has been speculated that variational algorithms of this type may be the first to demonstrate a quantum advantage over classical supercomputers for practical problems~\cite{Wecker:2015a}.

In this work, we explored the theory of a variational hybrid quantum-classical approach beyond its original context to more general problems.  We explored two potential candidates for an ansatz that may allow one to go beyond classical computation, namely a variational adiabatic formulation and the unitary coupled cluster method.  A simple connection between the second order unitary coupled cluster method and universal gate models of quantum computation was demonstrated.  Moreover, we showed that the variational formalism allows for a natural form of error suppression for some quantum problems in a pre-threshold device.  From a practical computational side, we showed that careful grouping of terms and truncation can offer significant cost savings in the use of this algorithm.  Finally we improved the classical subparts of the algorithm and found that advances in derivative free optimization offer dramatic cost savings over previous implementations.

Only time will tell if variational algorithms will be the first to surpass classical computers and if they can accomplish that feat on a pre-threshold device.  Regardless of this outcome, the variational framework offers a powerful perspective for the development of tools throughout quantum computation and the perspectives we have investigated and extended in this work will aid in this endeavor.
\subsection*{Acknowledgements}
R.B. thanks Alireza Shabani for helpful discussion about Pontryagin's Principle.  J.R.M. is supported by the Luis W. Alvarez fellowship in Computing Sciences at Lawrence Berkeley National Laboratory. J. R., R. B. and A. A.-G. acknowledge the Air Force Office of Scientific Research for support under Award: FA9550-12-1-0046. A. A.-G. acknowledges the Army Research Office under Award: W911NF-15-1-0256 and the Defense Security Science Engineering Fellowship managed by the Office of Naval Research.

\section{Appendix}
\subsection{Fermion commutation relations}
Here we document the generic commutation relations from the interacting fermion Hamiltonian without assuming anything about whether an index corresponds to an occupied or unoccupied orbital in a reference state.
\begin{align}
[a_i^\dagger a_j, a_p^\dagger a_q] &= a_i^\dagger a_q \delta_{pj} - a_p^\dagger a_j \delta_{iq} \\
[a_i^\dagger a_j, a_p^\dagger a_q^\dagger a_r a_s] &= a_i^\dagger a_q^\dagger a_r a_s \delta_{pj} - a_i^\dagger a_p^\dagger a_r a_s \delta_{jq} \\
& - a_p^\dagger a_q^\dagger a_r a_j \delta_{si} + a_p^\dagger a_q^\dagger a_s a_j \delta_{ri} \notag \\
[a_i^\dagger a_j^\dagger a_k a_l, a_p^\dagger a_q^\dagger a_r a_s] &= 
a_i^\dagger a_j^\dagger a_r a_s (\delta_{kq} \delta_{lp} -\delta_{kp}\delta_{lq}) \\
& - a_p^\dagger a_q^\dagger a_k a_l (\delta_{si} \delta_{rj} - \delta_{ri} \delta_{sj}) \notag  \\
& -a_i^\dagger a_j^\dagger a_q^\dagger a_k a_r a_s \delta_{lp} + a_p^\dagger a_q^\dagger a_i^\dagger a_r a_k a_l \delta_{si} \notag  \\
 &+ a_i^\dagger a_j^\dagger a_q^\dagger a_l a_r a_s \delta_{kp} 
- a_p^\dagger a_q^\dagger a_i^\dagger a_s a_k a_l \delta_{ri} \notag  \\
 &+a_i^\dagger a_j^\dagger a_p^\dagger a_k a_r a_s \delta_{lq}
- a_p^\dagger a_q^\dagger a_i^\dagger a_r a_k a_l \delta_{sj} \notag  \\
&- a_i^\dagger a_j^\dagger a_p^\dagger a_l a_r a_s \delta_{kq}
+ a_p^\dagger a_q^\dagger a_i^\dagger a_s a_k a_l \delta_{rj} \notag 
\end{align}
\subsection{Eigenvector Bound}
In this section we derive the bound on the quality of the eigenvector stated in the text as determined by the variance of the operator.  The ground state is different than general eigenstates in allowing a slightly easier derivation, so we split the derivations into two separate sub sections.
\subsubsection{Ground State}
Beginning with a calculation of the average energy in terms of the eigenvalues and weights of eigenvectors in a state $\ket{\Psi}$ decomposed into eigenvectors of $H$ as $\ket{\Psi}=\sum_i c_i \ket{\chi_i}$,
\begin{align}
\avg{H} &=  |c_1|^2 \lambda_1 + \sum_{i>1} |c_i|^2 \lambda_i \notag \\
&\geq |c_1|^2 \lambda_1 + \sum_{i>1} |c_i|^2(\lambda_1 + \Delta) \notag \\
&= |c_1|^2 \lambda_1 + (1- |c_1|^2)(\lambda_1 + \Delta) \notag \\
&= \lambda_1+\Delta - |c_1|^2 \Delta \notag \\
&\geq \left(\avg{H} - \sqrt{\Var(\vec{\theta})}\right) + \Delta - |c_1|^2 \Delta
\end{align}
where $\Delta$ is a lower bound on the gap between the ground and first excited eigenvalue.  Rearranging yields the desired bound on the overlap with the ground state
\begin{align}
|c_1|^2 \geq \frac{\Delta - \sqrt{\Var(\vec{\theta})}}{\Delta}
\end{align}
where the promise that the error is less than the gap, i.e. $\sqrt{\Var(\vec{\theta})} < \Delta$ guarantees a positive bound, and the overlap estimate converges to 1 as $\Var(\vec{\theta})$ is reduced to $0$.
\subsubsection{General States}
Starting with an expression for the variance of $H$ over a state $\ket{\Psi} = \sum_i c_i \ket{\chi_i}$ where $\ket{\chi_i}$ are eigenvectors of $H$ with eigenvalue $\lambda_i$, we have
\begin{align}
\Var[H] &= (H-E)^2 \ket{\Psi} \notag \\
&= \sum_{i \neq k} (\lambda_i - E)^2 |c_i|^2 + (\lambda_k - E)^2 |c_k|^2.
\end{align}
where $E=\avg{H}$.  Our goal is to bound the value of $|c_k|^2$ based on a measured variance of the state with respect to $H$, $\Var[H]$ and a known bound on the gap $\Delta$.  Let $\alpha=(\lambda_k - E)^2$, from here we see that
\begin{align}
\Var[H] \geq \left( \Delta + \sqrt{\Var[H]} \right)^2 (1-|c_k|^2) + \alpha |c_k|^2
\end{align}
rearranging to have an expression for $|c_k|^2$ and letting $\gamma = \left( \Delta + \sqrt{\Var[H]} \right)^2$, we have
\begin{align}
|c_k|^2 \geq \frac{\gamma - \Var[H]}{\gamma - \alpha}.
\end{align}
Following our assumptions on the gap and errors, we know that and $0 \leq \alpha \leq \Var[H] < \gamma$, from which it follows that
\begin{align}
|c_k|^2 \geq \frac{\gamma - \Var[H]}{\gamma}
\end{align}

\bibliographystyle{apsrev4-1}
\bibliography{VQETheory}

\end{document}